\documentclass{article}

\PassOptionsToPackage{numbers,sort&compress}{natbib}

\usepackage[preprint]{neurips_2026}

\usepackage[utf8]{inputenc} 
\usepackage[T1]{fontenc}    
\usepackage{hyperref}       
\usepackage{url}            
\usepackage{booktabs}       
\usepackage{amsfonts}       
\usepackage{nicefrac}       
\usepackage{pifont}
\usepackage{microtype}      
\usepackage{xcolor}         
\usepackage[table]{xcolor}
\usepackage{adjustbox}
\usepackage{multicol}
\usepackage{multirow}
\usepackage{wrapfig}
\usepackage{enumitem}
\usepackage{amsmath}
\usepackage[capitalize]{cleveref}
\crefname{section}{Sec.}{Secs.}
\Crefname{section}{Section}{Sections}
\Crefname{table}{Table}{Tables}
\crefname{table}{Tab.}{Tabs.}

\definecolor{deltared}{RGB}{192,0,0}

\definecolor{myred}{RGB}{238,92,81}
\definecolor{codegreen}{RGB}{1,153,0}
\definecolor{bg-blue}{RGB}{242, 249, 255}
\definecolor{bg-green}{RGB}{245, 252, 245} 
\definecolor{bg-pink}{RGB}{255, 240, 240}  

\title{\textsc{RepoMirage}: Probing Repository Context Reasoning in Code Agents with Perturbations}

\author{
  Hanyu Li$^{1*}$\quad
  Yichi Zhang$^{2}$\thanks{Equal contribution; alphabetical order. (mail to: \texttt{thestudent@bupt.edu.cn, zyc22@mails.tsinghua.edu.cn})}\quad
  Speed Zhu$^{3}$\quad
  Hang Su$^{2}$\quad
  Jun Zhu$^{2}$\quad
  Yinpeng Dong$^{2}$
  \\[3pt]
$^1$ Beijing University of Posts and Telecommunications \quad
$^2$ Tsinghua University \quad 
$^3$ Tencent
}

\begin{document}

\maketitle

\begin{abstract}
Code agents are currently having skillful performance on repository-level software engineering benchmarks, but it remains unclear whether success on end-to-end tasks such as issue resolution truly reflects \textit{repository context reasoning}, the ability to identify the task-relevant information across multiple files and reason over the relations among them. To investigate this question, we introduce \textsc{RepoMirage}, a two-stage evaluation suite built on SWE-Bench Verified that adopts perturbation as a diagnostic tool to increase the demand for context reasoning by transforming how the repository is exposed. First, \textsc{RepoMirage}-Perturb applies three types of semantics-preserving repository-level perturbations, revealing a clear performance drop when correct solving requires broader context access. \textsc{RepoMirage}-Extend further turns perturbation-targeted structural bottlenecks into explicit tasks beyond issue resolution, where the average performance declines from 66.8\% in the original setting to 25.3\%, indicating a significant deficiency in repository context reasoning. Further trajectory analysis reveals an exploration drift, where agents access broader repository context but fail to turn it into effective structure information. Motivated by this observation, we propose \textsc{RepoAnchor}, a structure-first prototype workflow that separates repository exploration from downstream problem solving, and show that explicit structural scaffolding yields notable gains. These results uncover an previously overlooked gap in repository context reasoning for code agents and suggest that stronger structure-aware methods are potential to improve them.

\end{abstract}

\section{Introduction}

\begin{figure}[t]
\centering
\includegraphics[width=1.0\columnwidth]{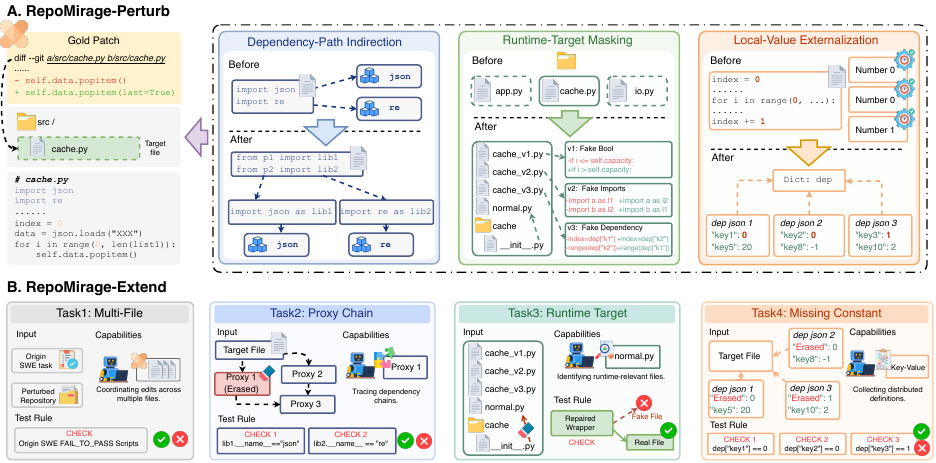}
\caption{\textbf{Overview of \textsc{RepoMirage}.}
\textsc{RepoMirage} is an evaluation suite for probing repository context reasoning in code agents.
\textbf{A. \textsc{RepoMirage}-Perturb.}
To test whether issue-resolution performance remains stable under higher repository-context demands, it applies three semantics-preserving perturbations while keeping the original task and evaluation unchanged.
\textbf{B. \textsc{RepoMirage}-Extend.}
To make repository context reasoning directly measurable, it builds four explicit task families on fully perturbed repositories by turning structural bottlenecks into task objectives.}
\label{fig:main}
\vspace{-3ex}
\end{figure}

Recent advances in large language models (LLMs) have extended code generation from function-level~\citep{du2024evaluating,chen2021codex} and single-file~\citep{jain2024livecodebench,hendrycksapps2021} settings to repository-level software engineering tasks~\citep{zan2025multiswebench,wu2025repomastereval,jimenez2024swebench,liu2023repobench}. As the deployment of LLMs becomes increasingly wide in practice, spanning diverse settings such as vibe coding~\citep{fawzy2025vibe} and cybersecurity analysis~\citep{xu2024large}, models are often required to operate over entire repositories rather than isolated code snippets~\citep{cui2026effects,zhang2024autocoderover}. Accomplishing these project-level applications essentially requires gathering distributed evidence across multiple files. Taking issue resolution as an example, the observed issue in one place may be caused by errors in another module, so determining a correct fix demands tracing calling relationships and execution constraints elsewhere in the codebase~\citep{xia2024agentless,yang2024swe}. Nevertheless, existing evaluations mostly treat such tasks in an end-to-end manner, without explicitly isolating the underlying ability to identify and organize task-relevant information across the code context~\citep{rashid2025swepolybench,Deng2025SWEBenchPC}. We refer to this capability as \emph{repository context reasoning}, which typically involves multi-file understanding and cross-file reasoning in a repository.

Despite the impressive progress of code agents on repository-level benchmarks, it remains unclear whether their increasing scores are a reliable indicator of repository context reasoning. We investigate this question through SWE-Bench Verified~\citep{chowdhury2024swebenchverified}, where our experiments reveal a localized pattern in issue resolution. Among successfully resolved cases, advanced agents frequently inspect only a few files (see~\cref{fig:open_file_dist}), which indicates that high success rates may not necessarily come from sufficient reasoning over broader repository context. At the same time, simply constructing new benchmarks with stronger multi-file demands would not address this challenge, because they are built on new environments with different task formulations~\citep{zhang2025swebenchgoeslive,yang2025swe}, making failures hard to attribute. Therefore, we need a controlled evaluation that probes repository context reasoning from the existing end-to-end benchmarks with the original task semantics and codebase instances unchanged, while selectively examining how much successful solutions depend on broader context access and cross-file reasoning.

In this paper, we use perturbation as a diagnostic tool to facilitate this controlled evaluation, which has been adopted to study whether model success relies on robust understanding or superficial cues~\citep{dong2018boosting,mccoy-etal-2019-right}. We introduce this methodology to construct \textsc{RepoMirage}, a perturbation-based evaluation suite organized as a two-stage evaluation, as shown in~\cref{fig:main}. We begin with \textsc{RepoMirage}-Perturb, which applies semantics-preserving perturbations to the task-relevant context in each benchmark instance while keeping the original task and evaluation protocol unaffected. This stage aims to reveal whether strong benchmark performance remains stable when successful resolution requires more repository context reasoning. Concretely, we design three perturbation strategies that obscure previously explicit dependency paths, hide true runtime targets behind structural indirection, and externalize locally available definitions into cross-file references, in order to force agents to read more files and reason over the context rather than following the localized pattern. Empirically, the resolved rates of eight leading models drop relatively by 27.0\% on average while reading substantially more files, showing that strong performance on issue-solving benchmarks does not directly translate to context reasoning.


Having established this notable performance gap under controlled perturbations, we then introduce \textsc{RepoMirage}-Extend to make repository context reasoning more explicitly measurable through new task formulations. Starting from the same perturbed instances, \textsc{RepoMirage}-Extend converts each perturbation strategy into a corresponding task objective, including proxy-chain completion, runtime target identification, and missing-constant recovery. These three tasks are combined with the original issue-resolution task involving multi-file gold patches. Derived from our perturbation strategies, these new tasks correspond to the fundamental structural relations practically common within project-level problems. On the same underlying benchmark instances, average performance across eight frontier models drops from 66.8\% under the original issue-resolution task to 25.3\% under our perturbation-derived task formulations, further indicating that successfully resolving an instance in the original benchmark does not imply the ability to understand and reason across the context in the repository. More importantly, the consistently low completion rates on these explicit tasks suggest that current code agents are still limited in repository context reasoning.


To better understand this deficiency, we delve into the behavioral patterns in agent trajectories under perturbations. We notice that while agents spend a larger proportion of their actions searching and reading files, it fails to result in effective solving. Instead, agents become increasingly prone to remain exploring rather than making editing decisions. This points to a mechanistic bottleneck where current agents can access a broader repository context, but often fail to organize the multi-file information into a coherent structural scaffold for task completion, which is validated by the improved performance when structural hints are provided. We then implement this idea into \textsc{RepoAnchor}, a prototype workflow that separates task-related structure exploration from downstream problem solving, by first constructing a summary of relevant repository context and then using it to guide subsequent actions. The resulting performance gains with \textsc{RepoAnchor} suggest that explicit structure understanding is a promising direction for improving repository context reasoning in code agents.

We summarize our main contributions as: \ding{172} \textbf{\textit{Conceptually}}: We identify an evaluation gap between strong performance on repository-level issue-resolution benchmarks and genuine repository context reasoning, showing that benchmark success does not necessarily arise from multi-file understanding and cross-file reasoning. \ding{173}
\textbf{\textit{Technically}}: We utilize perturbation as a diagostic tool for capability probing and present \textsc{RepoMirage}, a perturbation-based evaluation suite, to quantitatively measure this gap through two complementary settings, \textsc{RepoMirage}-Perturb and \textsc{RepoMirage}-Extend, revealing the deficiency in repository context reasoning for code agents. \ding{174}
\textbf{\textit{Mechanistically}}: We provide practical evidence that current agents suffer from exploration drift under higher repository-context demands, and show that structure-first scaffolding can substantially improve performance.

\section{Preliminary}

In this section, we describe our experimental setup and present an initial analysis of the file access behavior on repository-level issue resolution.

\subsection{Experimental Setup}

We study whether existing benchmarks on project-level problem solving can reliably reflect repository context reasoning in code agents. For controlled evaluation, we use SWE-Bench Verified as the seed benchmark, since it is a widely used human-validated benchmark built on real GitHub codebases for repository-level issue resolution in LLMs. We preserve the program semantics and executable functionality of each instance under perturbations using the provided test cases. All runs are conducted with \texttt{mini-swe-agent}~\citep{yang2024swe} under the standard bash-only setting, a simple but effective framework used in the SWE-Bench~\citep{jimenez2024swebench} leaderboard, which enables standardized comparison while minimizing confounding differences in agent design. We record agent trajectories, especially shell-command actions, for later analysis. We evaluate eight advanced LLMs, mainly using GPT-5~\citep{singh2025openai} and DeepSeek-V3.2~\citep{liu2025deepseek} for analysis. More details are provided in Appendix~\ref{sec:execution_protocol}.

\subsection{File Access Analysis in Repository-Level Issue Resolution}
\label{sec:file-access-analysis}

\begin{wrapfigure}[12]{r}{0.48\textwidth}
    \vspace{-23pt}
    \centering
    \setlength{\abovecaptionskip}{1pt}
    \includegraphics[width=\linewidth]{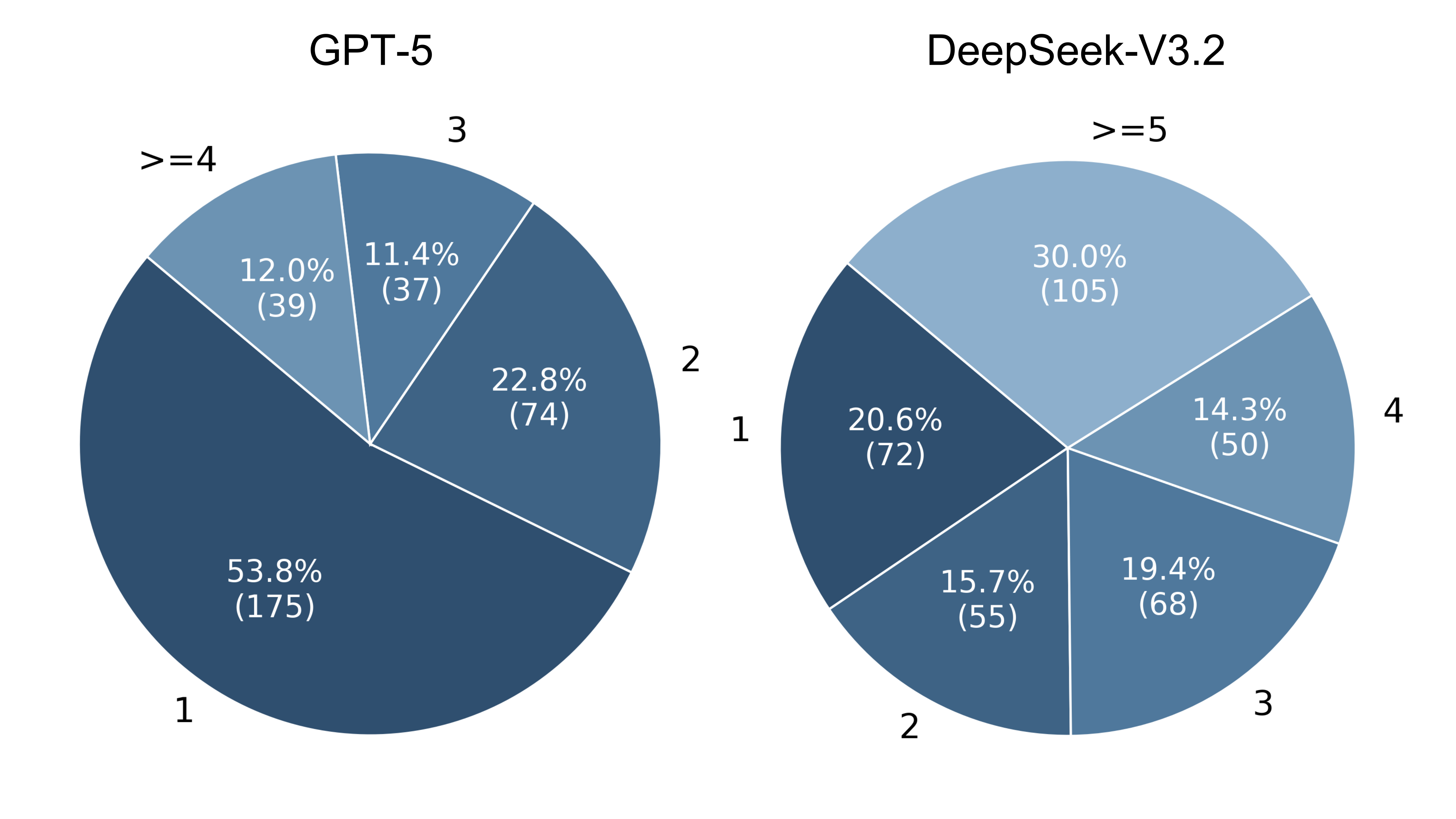}
    \vspace{-4ex}
    \caption{\textbf{Numbers of accessed files} per instance resolved in SWE-bench Verified for GPT-5 (avg. 2.15) and DeepSeek-V3.2 (avg. 3.97).}
    \label{fig:open_file_dist}
\end{wrapfigure}

While prior work shows that more than 80\% of SWE-Bench Verified instances ultimately require edits to only a single file~\citep{dihan2026patchrecall}, and that advanced models can identify a buggy file path with up to 76\% accuracy even without repository structure~\citep{liang2025swe}, these findings are descriptive and not grounded in the execution during issue resolution. We hereby directly examine how many files are actually accessed in successfully resolved SWE-Bench Verified instances. As shown in ~\cref{fig:open_file_dist}, file access is highly localized among resolved cases. For GPT-5, 53.8\% of resolved instances involve inspecting only one file, and 88.0\% involve no more than three files, while for DeepSeek-V3.2, 55.7\% of resolved instances still stay within three files. These results suggest that successful issue resolution is often achieved after accessing only a narrow portion of the repository. This raises the question of \emph{whether strong performance on existing repository-level benchmarks reliably reflects repository context reasoning in code agents, and, if not, how this capability can be measured more directly}.


\section{Uncovering Performance Gap Under Repository-Level Perturbations}

To probe repository context reasoning in code agents from the success in end-to-end issue-resolution benchmarks, we need a controlled evaluation to selectively isolate this capability while preserving the semantics and functionality of the original task. Perturbations provide a natural tool for this purpose. They have been widely used in vision and language tasks to study whether models rely on robust representations or superficial patterns by changing how input information is presented~\citep{ilyas2019adversarial,mccoy-etal-2019-right,ribeiro2020beyond}. In this paper, we introduce \textsc{RepoMirage}, a perturbation-based evaluation suite built on the issue-resolution benchmark, which uses targeted interventions to emphasize repository context reasoning while preserving the underlying codebase semantics and executable behavior of each instance.

In this section, we begin with \textsc{RepoMirage}-Perturb, which applies repository-level perturbation strategies to the original benchmark instances, and evaluate the performance of code agents on the same task of issue resolution, aiming to see whether there is a gap due to increasing context reasoning.




\subsection{Repository-Level Perturbation Strategies}

To realize the controlled evaluation, our repository-level perturbations are designed to make task-relevant information in the context less locally accessible, while keeping the instance semantically and functionally equivalent to the original setting. Then, performance changes can be interpreted more directly as evidence of limited repository context reasoning in current code agents. As illustrated in~\cref{fig:main}, we hereby propose the following repository-level perturbation strategies:
\begin{enumerate}[left=0pt]
    \item \textbf{Dependency-path indirection.} 
    Import statements often expose cross-file relations directly, allowing agents to localize relevant files without much structural inference over the context. To weaken this cue, we replace direct imports in target files with multi-hop proxy paths. Each imported library is rerouted through a constructed four-layer proxy chain, so that the visible import no longer directly reveals the true dependency. As a result, understanding the role of a referenced component requires tracing the dependency path across multiple files.

    \item \textbf{Runtime-target masking.}
    File names and paths often make the patch target easy to localize given issue descriptions, offering agents easier textual relevance as a proxy for runtime relevance. To weaken this cue, we rename the original target file, create a same-named module directory at its previous location, and use \texttt{\_\_init\_\_.py} to re-export the renamed file. We also place nearby fake files that remain globally similar to the target but contain locally perturbed details. As a result, the agent must distinguish files that merely appear relevant from the file that is actually used at runtime, requiring more cross-file comparison and reasoning.

    \item \textbf{Local-value externalization.}
    Task-relevant values such as constant definitions are often placed directly near the code using them. This local availability allows the agent to understand or modify the target logic within a single file, without understanding how program behavior depends on information stored elsewhere. To weaken this cue, we move local constant values from target files into external JSON resources and modify the code to load them at runtime. As a result, the agent must connect the local program logic with the external resource file and recover the value relationship across files, rather than relying on values that are directly accessible in the target file.

\end{enumerate}
To preserve comparability with the original benchmark, we verify that all perturbed instances retain their original semantics and functionality under the corresponding validation scripts. Full construction details and validation procedures are provided in the Appendix~\ref{sec:perturb_construction}.


\subsection{Code Agent Performance on \textsc{RepoMirage}-Perturb}

We apply three designed perturbations on the instances from SWE-Bench Verified and construct the dataset of \textsc{RepoMirage}-Perturb, on which we examine how code agents perform on the same task of issue resolution. We report both final resolved rates and the average numbers of accessed files.



\begin{table}[h!]
\vspace{-2ex}
\centering
\small
\caption{\textbf{Performance on \textsc{RepoMirage}-Perturb.} We compare model performance on the original SWE-Bench instances and their perturbed counterparts under the same issue task.}
\resizebox{\textwidth}{!}{
\begin{tabular}{l|ccc|ccc}
\toprule
\multirow{2}{*}{Model} 
& \multicolumn{3}{c|}{Resolved \%} 
& \multicolumn{3}{c}{Avg. \#Files} \\
\cmidrule(lr){2-4}\cmidrule(lr){5-7}
& SWE-Bench & Perturb & Drop 
& SWE-Bench & Perturb & Ratio \\
\midrule
GPT-4.1~\citep{openai2025gpt41}        
& 38.40 & 18.20 & \cellcolor{red!65}-52.60\%  
& 1.69 & 7.10 & \cellcolor{blue!45}4.20$\times$ \\

GPT-5~\citep{singh2025openai}          
& 65.00 & 49.00 & \cellcolor{red!32}-24.61\%  
& 2.83 & 7.04 & \cellcolor{blue!25}2.49$\times$ \\

Gemini\mbox{-}3.1\mbox{-}Pro~\citep{google2026gemini31}
& 70.60 & 54.40 & \cellcolor{red!30}-22.95\% 
& 17.32 & 24.94 & \cellcolor{blue!15}1.44$\times$ \\

Claude\mbox{-}Sonnet\mbox{-}4.6~\citep{claude-system-card} 
& 75.20 & 63.20 & \cellcolor{red!20}-15.96\% 
& 2.20 & 6.76 & \cellcolor{blue!35}3.07$\times$ \\

DeepSeek-V3.2~\citep{liu2025deepseek}  
& 70.00 & 52.00 & \cellcolor{red!33}-25.71\%  
& 4.66 & 14.32 & \cellcolor{blue!35}3.07$\times$ \\

MiniMax-M2.7~\citep{minimaxm27}   
& 78.20 & 65.40 & \cellcolor{red!20}-16.37\%  
& 2.24 & 6.92 & \cellcolor{blue!35}3.09$\times$ \\

Qwen3-Coder-Next~\citep{qwen_qwen3_coder_next_tech_report}
& 69.20 & 42.60 & \cellcolor{red!48}-38.44\% 
& 4.66 & 26.90 & \cellcolor{blue!60}5.77$\times$ \\

Qwen3.6-35B-A3B~\citep{qwen3.6-35b-a3b}
& 67.80 & 53.40 & \cellcolor{red!34}-21.24\%  
& 2.52 & 11.90 & \cellcolor{blue!45}4.72$\times$ \\
\bottomrule
\end{tabular}}
\label{tab:perturb_main}
\vspace{-3ex}
\end{table}

\cref{tab:perturb_main} compares model performance on the original \textsc{SWE-Bench Verified} instances and their \textsc{RepoMirage}-perturbed counterparts under the same issue-resolution task. Across all eight evaluated models, resolved rates consistently decrease under perturbation, with the average resolved rate dropping from \textbf{66.80\%} to \textbf{49.78\%} and relative drops ranging from \textbf{15.96\%} for Claude-Sonnet-4.6~\citep{claude-system-card} to \textbf{52.6\%} for GPT-4.1~\citep{openai2025gpt41}. This indicates that the success in issue resolution is not stable against the repository-level perturbations. Meanwhile, the perturbations substantially increase the amount of repository context that agents must access, with the average number of accessed files rising from \textbf{4.77} to \textbf{13.24}. This confirms that \textsc{RepoMirage}-Perturb successfully expands the codebase context necessary to solve the issues, further amplifying the importance of repository context reasoning in the task. Altogether, these results suggest that, even under the same issue-resolution task, increasing the demand for repository context reasoning leads to substantially worse performance, confirming that the gap between issue-resolution success and this capability is significant.

\begin{wrapfigure}[18]{r}{0.5\textwidth}
    \vspace{-10pt}
    \centering
    \setlength{\abovecaptionskip}{1pt}
    \includegraphics[width=\linewidth]{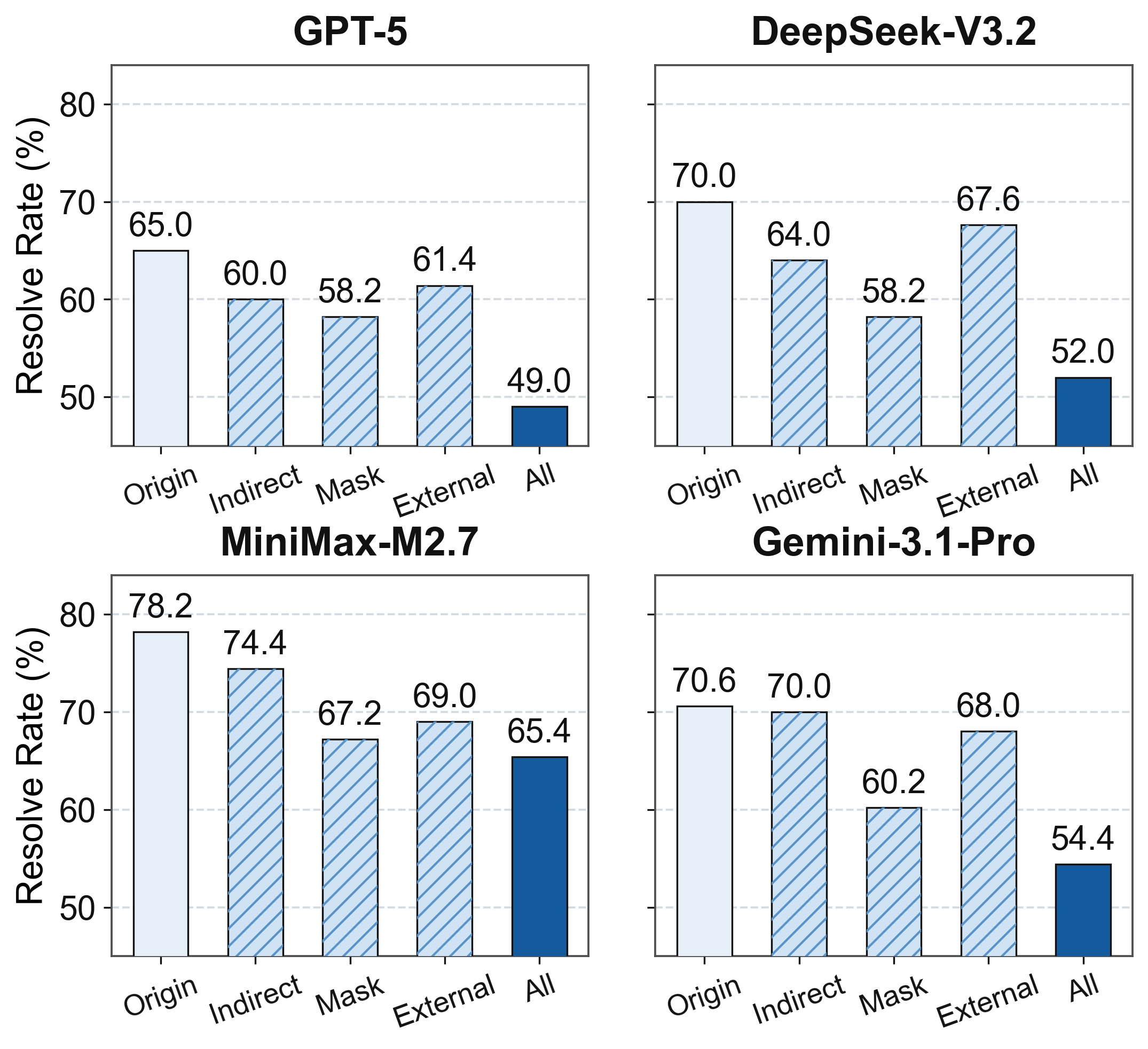}
    \vspace{-4ex}
    \caption{\textbf{Ablation study} of different repository perturbations across models.}
    \label{fig:perturb_ablation}
\end{wrapfigure}
We further examine whether the performance drop is driven by a single perturbation type or by their combined effect. \cref{fig:perturb_ablation} reports an ablation over dependency-path indirection, runtime-target masking, and local-value externalization. Each perturbation alone reduces performance relative to the baseline, indicating that agents benefit from all three types of direct local cues in standard issue resolution. Applying all perturbations together further lowers the resolved rates below any single-perturbation setting, showing that the three perturbations target complementary aspects of repository context reasoning.


Overall, these results suggest that standard issue-resolution tasks do not reliably reflect repository context reasoning in code agents. This motivates a more explicit probe of this capability by deriving new task formulations from the same instances to isolate the evaluation objective.
\section{Measuring Repository Context Reasoning Beyond Issue Resolution}

In this section, we move beyond issue resolution and introduce \textsc{RepoMirage}-Extend, a dataset built from the same perturbed instances as \textsc{RepoMirage}-Perturb, which are reformulated into new tasks derived from the perturbation strategies that explicitly probe the repository context reasoning.



\subsection{Perturbation-Derived Task Design}


The performance degradation under perturbations in the original issue-resolution task suggests that the structural relations targeted by \textsc{RepoMirage}-Perturb are key elements of repository context reasoning, which indeed correspond to recurring capabilities in practical software engineering, such as tracing dependency chains, identifying runtime-relevant files, collecting distributed definitions, and coordinating edits across multiple files. Based on this observation, we construct four task families in \textsc{RepoMirage}-Extend, with one of them inherited from the original setting while the other three derived directly from the corresponding perturbation strategies by turning each targeted structural bottleneck into an explicit task objective, which are illustrated in~\cref{fig:main} and introduced below:
\begin{enumerate}[left=0pt]
    \item \textbf{Multi-File Issue Resolution.} We notice that a part of the original benchmark already involves gold patches spanning multiple files. Such cases naturally require stronger coordination across editing locations and therefore place higher demands on repository context reasoning. We retain these instances and evaluate how agents perform on them under perturbations, asking whether they can still reason over the context and produce multi-file edits that pass the official test cases.
    \item \textbf{Proxy Chain Completion.} Dependency-path indirection weakens the direct visibility of cross-file dependency relations, so agents can no longer recover the true imported component from local import statements alone. This places greater demands on tracing dependency chains across files. To make this demand explicit, we construct a proxy-chain completion task by erasing one intermediate import in the perturbed chain. The agent must reconstruct the missing proxy implementation so that the target file resolves to the correct underlying dependency. A task is considered solved if the restored chain passes dependency-specific checks and recovers the expected imported objects.
    \item \textbf{Runtime Target Identification.} Runtime-target masking weakens the direct correspondence between file-level surface cues and the file that is actually executed at runtime. As a result, successful solving requires distinguishing merely plausible files from the true runtime-relevant target. Starting from a \textsc{RepoMirage}-Perturb repository, we conceal the reference inside the synthetic module that re-exports the true target file. The agent must inspect the wrapper module, compare the candidate files, and reconstruct the mapping from the synthetic module to the actual runtime target. This task is considered solved if the agent identifies the correct runtime file specified by the perturbed wrapper structure.
    \item \textbf{Missing-Constant Recovery.} Local-value externalization moves task-relevant values from the target file into external resources, requiring the agent to recover value associations across files rather than from local context alone. To evaluate this ability more explicitly, we remove the keys of several key-value pairs in the external JSON files while keeping their values unchanged. The agent must inspect the target code, identify where the missing entries are used, infer the role of each key from the local program logic, and reconstruct the corresponding key-value mapping. A task is considered solved if the recovered keys match the original externalized definitions.
\end{enumerate}

\begin{wrapfigure}[10]{r}{0.28\textwidth}
    \vspace{-20pt}
    \centering
    \setlength{\abovecaptionskip}{1pt}
    \includegraphics[width=\linewidth]{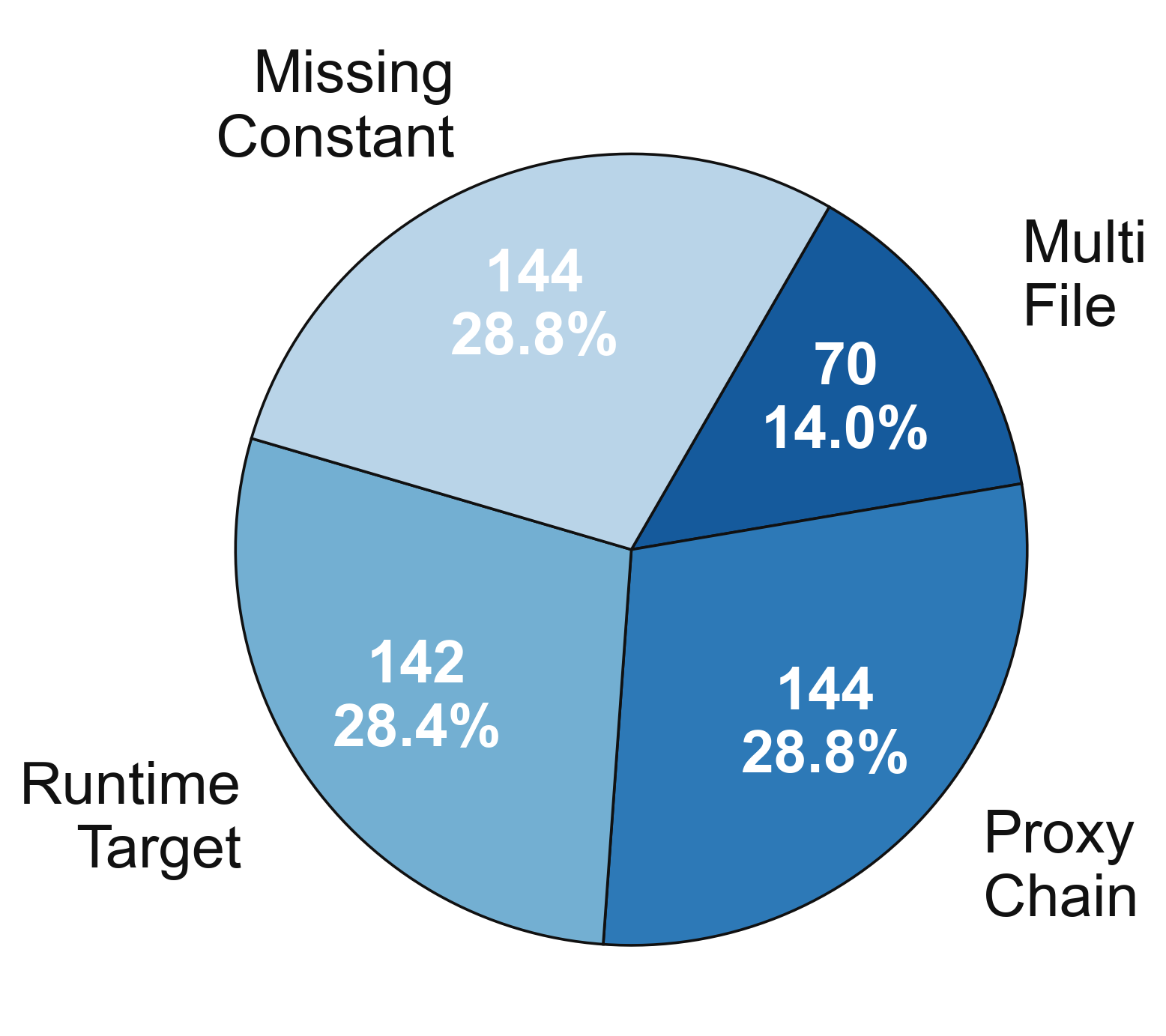}
    \caption{\textbf{Task distribution} in \textsc{RepoMirage}-Extend.}
    \label{fig:task_metadata}
\end{wrapfigure}
Instead of converting each benchmark instance into all four tasks, we assign every original instance to exactly one task family according to the structural signal it most naturally instantiates. Instances with multi-file gold patches are retained for Multi-File Issue Resolution, while the remaining instances are distributed across the three perturbation-derived tasks in a structure-aware manner based on the most salient structural signal for each instance. The task distribution is controlled to be roughly balanced and the distribution is displayed in~\cref{fig:task_metadata}. Full partition details are provided in the Appendix~\ref{sec:task_construction}.

\subsection{Code Agent Performance on \textsc{RepoMirage}-Extend}

We evaluate code agents instantiated with different frontier models on the four tasks in \textsc{RepoMirage}-Extend. \cref{tab:extend_results} reports the success rates on each task. Overall, \textsc{RepoMirage}-Extend substantially reduces agent performance across models and task types compared to the original task of issue resolution on the same instances. Averaged across the models, the success rate over all instances drops from \textbf{66.80\%} in the original settings to \textbf{25.25\%} on \textsc{RepoMirage}-Extend, showing that the deficiency of current agents in recovering the repository context through in-depth reasoning. Moreover, performance in the original setting does not show an exact correspondence with performance on new tasks. For instance MiniMax-M2.7 performs strongly on the original tasks but drops sharply on \textsc{RepoMirage}-Extend. This mismatch indicates that standard issue-resolution benchmarks do not fully capture repository context reasoning, and that \textsc{RepoMirage}-Extend measures a distinct capability beyond producing a passing patch.

The difficulty also differs across task categories. Proxy Chain Recovery is the most challenging on average, with mean success dropping to \textbf{17.19\%}, indicating that tracing dependency paths across files remains difficult even for strong agents. Multi-File Issue Resolution is similarly challenging (\textbf{17.86\%}), reflecting the difficulty of coordinating realistic multi-file edits under weakened repository cues. Runtime Target Identification and Missing Constant Recovery show larger variation across models, with average success rates of \textbf{28.26\%} and \textbf{33.94\%}, respectively, suggesting that runtime-target localization and cross-file value association are not uniformly reliable across current agents.

Overall, these results show that standard issue-resolution benchmarks do not fully reflect repository context reasoning. When this capability is made more explicit through \textsc{RepoMirage}-Extend rather than being left implicit in end-to-end patch solving, current frontier models still perform far from satisfaction, indicating that repository context reasoning remains a clear limitation of code agents.

\begin{table}[t]
\centering
\vspace{-2ex}
\caption{\textbf{Performance on \textsc{RepoMirage}-Extend.}
\textit{Org.} denotes performance on the original SWE-Bench Verified for the same instance subset, and \textit{Extend} denotes performance on \textsc{RepoMirage}-Extend. \textbf{Bold} marks the best result, and \underline{underlining} marks the second-best result.}
\resizebox{\textwidth}{!}{
\begin{tabular}{l|cc|cc|cc|cc|l}
\toprule
\multirow{2}{*}{Model} & \multicolumn{2}{|c}{Multi-File} & \multicolumn{2}{|c}{Proxy Chain}& \multicolumn{2}{|c}{Runtime Target}& \multicolumn{2}{|c|}{Missing Constant} & \multirow{2}{*}{Avg.}  \\
\cmidrule{2-3}\cmidrule{4-5}\cmidrule{6-7}\cmidrule{8-9}
& Org. & Extend & Org. & Extend & Org. & Extend & Org. & Extend \\ 
\midrule
GPT-4.1~\citep{openai2025gpt41} & 10.00 & 2.86 & 45.14 & 4.17 & 40.14 & 3.52 & 43.75 & 2.78 &  3.40   \\
GPT-5~\citep{singh2025openai}    & 31.43 & \textbf{25.71} & 70.83 & 11.11 & 68.31 & 35.92 & 72.22 & 36.80 &  27.60   \\
Gemini\mbox{-}3.1\mbox{-}Pro~\citep{google2026gemini31} & 38.57 & 14.29 & 77.08 & 22.22 & 76.06 & \underline{43.66} & 74.31 & \textbf{71.53} &   \textbf{41.40} \\
Claude\mbox{-}Sonnet\mbox{-}4.6~\citep{claude-system-card} & 41.43 & 15.71 & 80.56 & \textbf{36.81} & 80.28 & 30.28 & 81.25 & 23.61 & 28.20 \\
DeepSeek-V3.2~\citep{liu2025deepseek} & 35.71 & 20.00 & 74.31 & 11.80 & 75.35 & \textbf{59.86} & 77.08 & \underline{57.64} &   \underline{39.80} \\
MiniMax-M2.7~\citep{minimaxm27} & 45.71 & 22.86 & 81.25 & 7.64 & 86.62 & 26.76 & 81.94 & 6.25 &   14.80 \\
Qwen3-Coder-Next~\citep{qwen_qwen3_coder_next_tech_report} & 44.29 & 17.14 & 72.92 & \underline{29.17} & 75.35 & 7.04 & 71.53 & 34.03 &   22.60 \\
Qwen3.6-35B-A3B~\citep{qwen3.6-35b-a3b} & 31.43  & \underline{24.29} & 80.56 & 14.58 & 72.41 & 19.01 & 68.00 & 38.89 &   24.20  \\
\bottomrule
\end{tabular}}
\label{tab:extend_results}
\vspace{-2ex}
\end{table}






\section{Diagnosing and Mitigating Failures in Repository Context Reasoning}

In this section, we diagnose these failures through the lens of repository exploration. We first analyze agent trajectories in \textsc{RepoMirage} to examine how behavior changes under perturbations. We then test whether structural hints improve performance. Motivated by this, we introduce \textsc{RepoAnchor}, a prototype that prioritizes structure exploration as a potential mitigation.

\begin{figure*}[h!]
  \centering
  \vspace{-1ex}
  \includegraphics[width=1.0 \columnwidth]{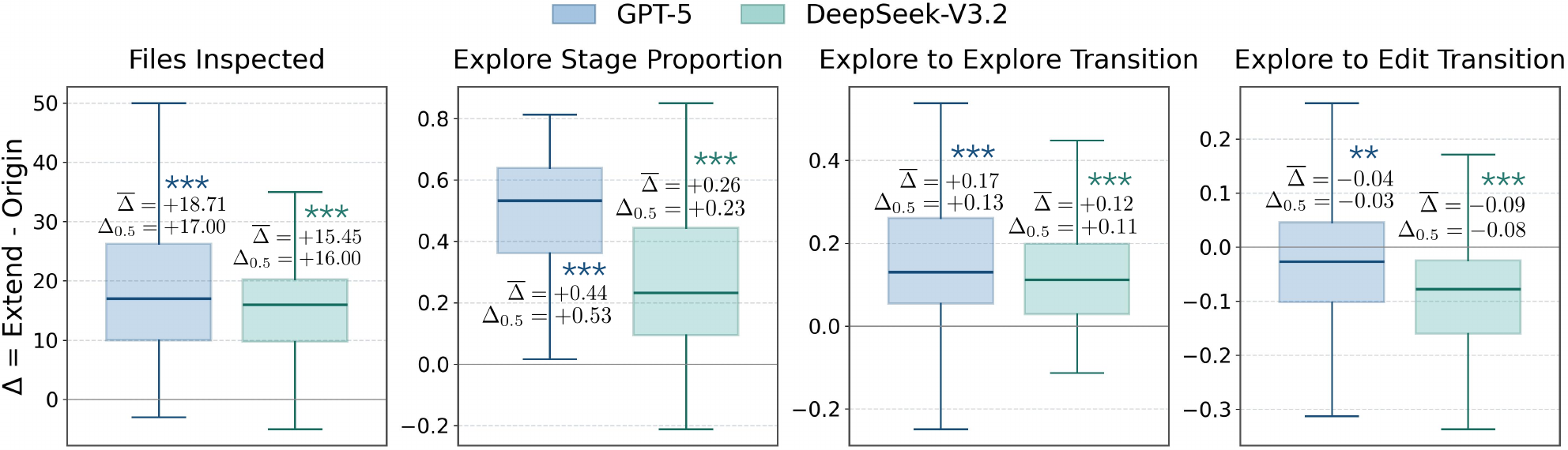}
  \vspace{-2ex}
  \caption{\textbf{Behavior shifts under \textsc{RepoMirage}-Extend.} $\Delta$ measures the change from SWE-Bench Verified to \textsc{RepoMirage}-Extend. Files Inspected counts distinct opened files; Explore Stage Proportion is the pre-edit step ratio; transition metrics report action-transition changes. $\bar{\Delta}$ and $\Delta_{0.5}$ denote the mean and median, respectively.}
  \label{fig:analysis_metric_deltas}
  \vspace{-2ex}
\end{figure*}

\subsection{Behavioral Diagnosis under Stronger Context Demands}
\label{sec:exploration-metrics}








To understand why agents fail under stronger repository context demands, we analyze their action trajectories, similar to previous work~\citep{bouzenia2025understanding,ceka2025understanding,kim2026trajeval}. We group agent operations into \{\textit{explore}, \textit{edit}, \textit{test}\} by the specific shell commands. For each trajectory, we define the transition probability from actions as
\[
p_{a\rightarrow b}
=
\frac{\#(a\rightarrow b)}
{\sum_{s\in \mathcal{S}}\#(a\rightarrow s)},
\quad a,b\in\{\textit{explore},\textit{edit},\textit{test}\}.
\]
We summarize the differences compared to original issue resolution in the number of files explored and the steps of exploration in trajectories, as well as the action transition probabilities, to examine the behavioral patterns demonstrated by different agents during execution. Due to space limit, we report the analysis on \textsc{RepoMirage}-Extend in the main text.

As shown in~\cref{fig:analysis_metric_deltas}, the perturbations and new tasks from \textsc{RepoMirage} consistently encourage agents toward broader context exploration. For both GPT-5 and DeepSeek-V3.2, the number of files explored before editing increases significantly and the pre-edit exploration stage becomes longer, indicating that agents need to search and read more broadly before taking action. At the same time, the transition probability from exploration to exploration increases, while the transition from exploration to editing decreases. This means that the additional exploration does not effectively translate into concrete editing decisions and agents frequently remain in the exploration stage. Together, this phenomenon of exploration drift suggests that while stronger demands on repository context reasoning trigger broader repository access, code agents still struggle to convert the collected multi-file information into effective problem solving. As discussed in Appendix~\ref{sec:traj_on_perturb}, we observe the same trend on \textsc{RepoMirage}-Perturb, and finer-grained task-level analyses further confirm the stability of this conclusion while also showing that different task families induce different types of behavioral shift.

\subsection{Validating the Bottleneck with Structural Hints}

\begin{wraptable}[10]{r}{0.6\textwidth}
\vspace{-20pt}
\centering
\caption{\textbf{Effect of structural hints.} Providing repository-structure hints substantially improves performance on \textsc{RepoMirage}-Extend.}
\label{tab:structural_hints}
\vspace{4pt}
\small
\setlength{\tabcolsep}{3pt}
\renewcommand{\arraystretch}{0.92}

\begin{tabular}{@{}lccccc@{}}
\toprule
Model & Multi & Proxy & Runtime & Const. & Avg. \\
\midrule
GPT-5~\citep{singh2025openai}
& 25.71 & 11.11 & 35.92 & 36.80 & 27.60 \\
\rowcolor{bg-blue}
+ Hints
& 28.57 & 81.94 & 60.56 & 61.81 & 62.60 \\
\midrule
DeepSeek-V3.2~\citep{liu2025deepseek}
& 20.00 & 11.80 & 59.86 & 57.64 & 39.80 \\
\rowcolor{bg-blue}
+ Hints
& 24.29 & 86.11 & 84.51 & 85.42 & 76.80 \\
\bottomrule
\end{tabular}
\vspace{-8pt}
\end{wraptable}
The trajectory patterns above suggest a possible explanation that although agents explore more files, they may still fail to organize them into a usable structural understanding for downstream solving. To test this, we conduct a simple structural-hint intervention. Instead of requiring the agent to infer the perturbed repository structure entirely from interaction, we provide a compact summary describing the perturbation mechanism, the roles of the perturbed files, and the key structural relations needed for solving the task. These hints approximate the ideal outcome of repository exploration by directly supplying a structured view of the repository. As shown in~\cref{tab:structural_hints}, this intervention substantially improves performance, especially on the perturbation-derived tasks. This supports our interpretation that the main bottleneck is the failure to transform observed files into an actionable structural guidance. Full details of structural hints are provided in the Appendix~\ref{sec:hint_details}.



\subsection{\textsc{RepoAnchor}: Anchoring Code Agents by Prioritizing Repository Exploration}

\begin{figure*}[t]
\vspace{-2ex}
  \centering
  \includegraphics[width=1.0 \columnwidth]{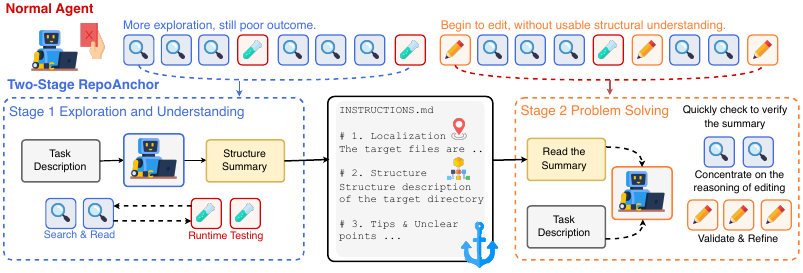}
  \vspace{-4ex}
  \caption{\textbf{Pipeline of \textsc{RepoAnchor}.}
A normal agent mixes its actions in one trajectory, failing to retrieve usable information. \textsc{RepoAnchor} separates the process into structure understanding and problem solving, using an intermediate \texttt{INSTRUCTIONS.md} file to pass repository context forward.}
\vspace{-2ex}
  \label{fig:repoanchor}
\end{figure*}







\begin{figure*}[t]
  \centering
  \vspace{-15pt}
  \includegraphics[width=1.0 \columnwidth]{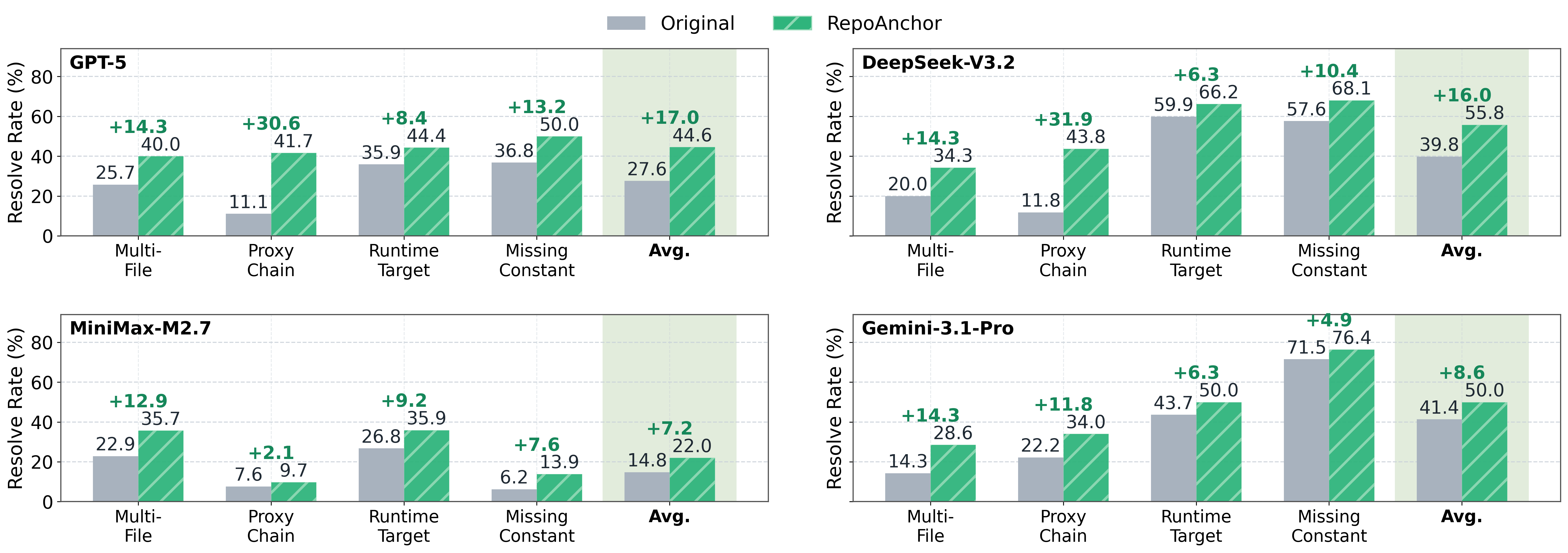}
  \vspace{-4ex}
  \caption{\textbf{Performance gains from \textsc{RepoAnchor}.} Resolved rates of four representative models before and after applying \textsc{RepoAnchor}. The consistent gains across task types and models show that structure-first repository understanding can improve downstream task solving.}
  \label{fig:repoanchor_results}
  \vspace{-15pt}
\end{figure*}

Motivated by the validation above, we introduce \textsc{RepoAnchor}, a initial mitigation method that separates structure exploration from downstream code construction. As shown in~\cref{fig:repoanchor}, instead of interleaving search, reading, editing, and testing within a single trajectory, \textsc{RepoAnchor} organizes the process into two stages. In the first stage, the agent is instructed to examine the codebase and write a structured summary into an intermediate Markdown file. This summary is expected to record the extracted repository context, including potential editing locations, runtime relations, relevant files, and other salient cross-file dependencies. In the second stage, the agent takes this Markdown summary together with the task description, and then focuses on implementation, editing, and patch generation. With this interface to transfer repository understanding between stages, \textsc{RepoAnchor} anchors downstream editing on a coherent structural view rather than on fragmented observations.



\cref{fig:repoanchor_results} shows that \textsc{RepoAnchor} consistently improves performance across four evaluated models on all tasks in \textsc{RepoMirage}-Extend, indicating that structure exploration provides a generally useful scaffold rather than benefiting only a specific model. Notably, the gains on Multi-File Issue Resolution are greater than those given the hints, rendering the informativeness of the structure summary. This suggests that explicit structure anchoring is a promising direction for improving code agents under complex repository settings demanding context reasoning.

\section{Related Work}

\subsection{Code Agent Benchmarks}

Code evaluation has progressed from localized coding to repository-level software engineering. Early benchmarks target function-level completion~\citep{chen2021codex,austin2021program,evalplus} and single-file program generation~\citep{hendrycksapps2021,Li_2022,jain2024livecodebench}, while later work expands to class-level~\citep{du2024evaluating}, data-science~\citep{Lai2022DS1000}, and execution reasoning~\citep{gu2024cruxeval}. Recent benchmarks further evaluate repository-level tasks~\citep{zhang2023repocoder,liu2023repobench}, cross-file reasoning~\citep{ding2023crosscodeeval}, and real-repository settings~\citep{wu2025repomastereval}. SWE-bench~\citep{jimenez2024swebench,chowdhury2024swebenchverified} requires agents to fix real issues, and later benchmarks extend this line with fresher tasks~\citep{zhang2025swebenchgoeslive}, broader ecosystems~\citep{zan2025multiswebench,wang2025swebenchpp}, and harder engineering~\citep{rashid2025swepolybench,Deng2025SWEBenchPC}. Despite these advances, existing benchmarks still primarily measure end-to-end success, so \textsc{RepoMirage} weakens superficial cues to make repository context itself the object of evaluation.

\subsection{Evaluation with Perturbations}

Perturbation-based evaluation tests model reliability under input changes~\citep{10.5555/3495724.3497285}, with early robustness studies showing neural models' sensitivity to transformed inputs~\citep{szegedy2014intriguingproperties,goodfellow2015explainingharness}. In NLP, such benchmarks reveal brittle heuristics behind high accuracy~\citep{ribeiro-etal-2020-beyond,gardner-etal-2020-evaluating,mccoy-etal-2019-right}. Similar ideas have been applied to code generation. Prompt variations can change coding performance~\citep{wang-etal-2023-recode}, while code transformations such as variable renaming~\citep{variablerenaming2025} test reliance on surface code patterns rather than program logic~\citep{li2023cctesttestingrepairingcode,na-etal-2023-dip,orvalho2025largelanguagemodelsrobust}. Recent work further studies code reasoning~\citep{maveli2025largelanguagemodelscapture} and codebase-level understanding~\citep{lam2025codecrash,lee2025gistifycodebaselevelunderstandingruntime} under altered inputs. Different from the study concerning robustness, we take perturbation as a diagnostic tool to probe the certain capability of repository context reasoning.

\section{Conclusion}

In this paper, we aim to probe repository context reasoning in code agents from existing end-to-end benchmarks like SWE-Agent Verified. Using perturbation as a diagnostic tool, we introduce \textsc{RepoMirage}, a perturbation-based evaluation containing two stages. First, we develop \textsc{RepoMirage}-Perturb by applying three perturbation strategies to change how task-relevant information is exposed in the context and find that frontier agents degrade substantially on issue resolution with perturbed repository context. Based on it, we turn these perturbation-targeted bottlenecks into explicit tasks and construct \textsc{RepoMirage}-Extend, where the significant lower performance renders the deficiency in this capability. Our trajectory analysis further suggests that agents often fail to organize evidences into actionable structural understanding, which motivates \textsc{RepoAnchor}, a structure-first scaffolding workflow, to improve the performance and provide a potential solution.


\clearpage{}

\bibliographystyle{unsrtnat}
\bibliography{reference}

\begin{thebibliography}{58}
\providecommand{\natexlab}[1]{#1}
\providecommand{\url}[1]{\texttt{#1}}
\expandafter\ifx\csname urlstyle\endcsname\relax
  \providecommand{\doi}[1]{doi: #1}\else
  \providecommand{\doi}{doi: \begingroup \urlstyle{rm}\Url}\fi

\bibitem[Du et~al.(2024)Du, Liu, Wang, Wang, Liu, Chen, Feng, Sha, Peng, and Lou]{du2024evaluating}
Xueying Du, Mingwei Liu, Kaixin Wang, Hanlin Wang, Junwei Liu, Yixuan Chen, Jiayi Feng, Chaofeng Sha, Xin Peng, and Yiling Lou.
\newblock Evaluating large language models in class-level code generation.
\newblock In \emph{Proceedings of the IEEE/ACM 46th International Conference on Software Engineering}, pages 1--13, 2024.

\bibitem[Chen et~al.(2021)Chen, Tworek, Jun, Yuan, de~Oliveira~Pinto, Kaplan, Edwards, Burda, Joseph, Brockman, Ray, Puri, Krueger, Petrov, et~al.]{chen2021codex}
Mark Chen, Jerry Tworek, Heewoo Jun, Qiming Yuan, Henrique~Ponde de~Oliveira~Pinto, Jared Kaplan, Harri Edwards, Yuri Burda, Nicholas Joseph, Greg Brockman, Alex Ray, Raul Puri, Gretchen Krueger, Michael Petrov, et~al.
\newblock Evaluating large language models trained on code.
\newblock \emph{arXiv preprint arXiv:2107.03374}, 2021.

\bibitem[Jain et~al.(2024)Jain, Han, Gu, Li, Yan, Zhang, Wang, Solar-Lezama, Sen, and Stoica]{jain2024livecodebench}
Naman Jain, King Han, Alex Gu, Wen-Ding Li, Fanjia Yan, Tianjun Zhang, Sida Wang, Armando Solar-Lezama, Koushik Sen, and Ion Stoica.
\newblock Livecodebench: Holistic and contamination free evaluation of large language models for code.
\newblock \emph{arXiv preprint arXiv:2403.07974}, 2024.

\bibitem[Hendrycks et~al.(2021)Hendrycks, Basart, Kadavath, Mazeika, Arora, Guo, Burns, Puranik, He, Song, and Steinhardt]{hendrycksapps2021}
Dan Hendrycks, Steven Basart, Saurav Kadavath, Mantas Mazeika, Akul Arora, Ethan Guo, Collin Burns, Samir Puranik, Horace He, Dawn Song, and Jacob Steinhardt.
\newblock Measuring coding challenge competence with apps.
\newblock \emph{NeurIPS}, 2021.

\bibitem[Zan et~al.(2025)Zan, Huang, Liu, Chen, Zhang, Xin, Chen, Liu, Zhong, Li, et~al.]{zan2025multiswebench}
Daoguang Zan, Zhirong Huang, Wei Liu, Hanwu Chen, Linhao Zhang, Shulin Xin, Lu~Chen, Qi~Liu, Xiaojian Zhong, Aoyan Li, et~al.
\newblock Multi-swe-bench: A multilingual benchmark for issue resolving, 2025.
\newblock \emph{arxiv preprint arXiv:2504.02605}, 2025.

\bibitem[Wu et~al.(2025)Wu, Peng, Gao, Hu, Gan, Jiang, Tang, Deng, Guan, Gao, et~al.]{wu2025repomastereval}
Qinyun Wu, Chao Peng, Pengfei Gao, Ruida Hu, Haoyu Gan, Bo~Jiang, Jinhe Tang, Zhiwen Deng, Zhanming Guan, Cuiyun Gao, et~al.
\newblock Repomastereval: Evaluating code completion via real-world repositories.
\newblock In \emph{2025 40th IEEE/ACM International Conference on Automated Software Engineering (ASE)}, pages 3672--3683. IEEE, 2025.

\bibitem[Jimenez et~al.(2024)Jimenez, Yang, Wettig, Yao, Pei, Press, and Narasimhan]{jimenez2024swebench}
Carlos~E Jimenez, John Yang, Alexander Wettig, Shunyu Yao, Kexin Pei, Ofir Press, and Karthik~R Narasimhan.
\newblock {SWE}-bench: Can language models resolve real-world github issues?
\newblock In \emph{The Twelfth International Conference on Learning Representations}, 2024.

\bibitem[Liu et~al.(2023{\natexlab{a}})Liu, Xu, and McAuley]{liu2023repobench}
Tianyang Liu, Canwen Xu, and Julian McAuley.
\newblock Repobench: Benchmarking repository-level code auto-completion systems.
\newblock In \emph{The Twelfth International Conference on Learning Representations}, 2023{\natexlab{a}}.

\bibitem[Fawzy et~al.(2025)Fawzy, Tahir, and Blincoe]{fawzy2025vibe}
Ahmed Fawzy, Amjed Tahir, and Kelly Blincoe.
\newblock Vibe coding in practice: Motivations, challenges, and a future outlook -- a grey literature review.
\newblock \emph{arXiv preprint arXiv:2510.00328}, 2025.

\bibitem[Xu et~al.(2024)Xu, Wang, Li, Wang, Zhao, Chen, Yu, Liu, and Wang]{xu2024large}
HanXiang Xu, ShenAo Wang, Ningke Li, Kailong Wang, Yanjie Zhao, Kai Chen, Ting Yu, Yang Liu, and HaoYu Wang.
\newblock Large language models for cyber security: A systematic literature review.
\newblock \emph{ACM Transactions on Software Engineering and Methodology}, 2024.

\bibitem[Cui et~al.(2026)Cui, Demirer, Jaffe, Musolff, Peng, and Salz]{cui2026effects}
Kevin~Zheyuan Cui, Mert Demirer, Sonia Jaffe, Leon Musolff, Sida Peng, and Tobias Salz.
\newblock The effects of generative ai on high-skilled work: Evidence from three field experiments with software developers.
\newblock \emph{Management Science}, 2026.

\bibitem[Zhang et~al.(2024)Zhang, Ruan, Fan, and Roychoudhury]{zhang2024autocoderover}
Yuntong Zhang, Haifeng Ruan, Zhiyu Fan, and Abhik Roychoudhury.
\newblock Autocoderover: Autonomous program improvement.
\newblock In \emph{Proceedings of the 33rd ACM SIGSOFT International Symposium on Software Testing and Analysis}, pages 1592--1604, 2024.

\bibitem[Xia et~al.(2024)Xia, Deng, Dunn, and Zhang]{xia2024agentless}
Chunqiu~Steven Xia, Yinlin Deng, Soren Dunn, and Lingming Zhang.
\newblock Agentless: Demystifying llm-based software engineering agents.
\newblock \emph{arXiv preprint arXiv:2407.01489}, 2024.

\bibitem[Yang et~al.(2024)Yang, Jimenez, Wettig, Lieret, Yao, Narasimhan, and Press]{yang2024swe}
John Yang, Carlos~E Jimenez, Alexander Wettig, Kilian Lieret, Shunyu Yao, Karthik Narasimhan, and Ofir Press.
\newblock Swe-agent: Agent-computer interfaces enable automated software engineering.
\newblock \emph{Advances in Neural Information Processing Systems}, 37:\penalty0 50528--50652, 2024.

\bibitem[Rashid et~al.(2025)Rashid, Bock, Zhuang, Buchholz, Esler, Valentin, Franceschi, Wistuba, Sivaprasad, Kim, et~al.]{rashid2025swepolybench}
Muhammad~Shihab Rashid, Christian Bock, Yuan Zhuang, Alexander Buchholz, Tim Esler, Simon Valentin, Luca Franceschi, Martin Wistuba, Prabhu~Teja Sivaprasad, Woo~Jung Kim, et~al.
\newblock Swe-polybench: A multi-language benchmark for repository level evaluation of coding agents.
\newblock \emph{arXiv preprint arXiv:2504.08703}, 2025.

\bibitem[Deng et~al.(2025)Deng, Da, Pan, Yiming~He, Ide, Garg, Lauffer, Park, Pasari, Rane, et~al.]{Deng2025SWEBenchPC}
Xiang Deng, Jeff Da, Edwin Pan, Yannis Yiming~He, Charles Ide, Kanak Garg, Niklas Lauffer, Andrew Park, Nitin Pasari, Chetan Rane, et~al.
\newblock Swe-bench pro: Can ai agents solve long-horizon software engineering tasks?
\newblock \emph{arXiv preprint arXiv:2509.16941}, 2025.

\bibitem[Chowdhury et~al.(2024)Chowdhury, Aung, Shern, Jaffe, Sherburn, Starace, Mays, Dias, Aljubeh, Glaese, Jimenez, Yang, Ho, Patwardhan, Liu, and Madry]{chowdhury2024swebenchverified}
Neil Chowdhury, James Aung, Chan~Jun Shern, Oliver Jaffe, Dane Sherburn, Giulio Starace, Evan Mays, Rachel Dias, Marwan Aljubeh, Mia Glaese, Carlos~E. Jimenez, John Yang, Leyton Ho, Tejal Patwardhan, Kevin Liu, and Aleksander Madry.
\newblock Introducing {SWE}-bench verified, 2024.
\newblock URL \url{https://openai.com/index/introducing-swe-bench-verified/}.

\bibitem[Zhang et~al.(2025)Zhang, He, Zhang, Kang, Li, Xie, Wang, Wang, Huang, Fu, Nallipogu, Lin, Dang, Rajmohan, and Zhang]{zhang2025swebenchgoeslive}
Linghao Zhang, Shilin He, Chaoyun Zhang, Yu~Kang, Bowen Li, Chengxing Xie, Junhao Wang, Maoquan Wang, Yufan Huang, Shengyu Fu, Elsie Nallipogu, Qingwei Lin, Yingnong Dang, Saravan Rajmohan, and Dongmei Zhang.
\newblock Swe-bench goes live!
\newblock \emph{arXiv preprint arXiv:2505.23419}, 2025.

\bibitem[Yang et~al.(2025)Yang, Lieret, Jimenez, Wettig, Khandpur, Zhang, Hui, Press, Schmidt, and Yang]{yang2025swe}
John Yang, Kilian Lieret, Carlos~E Jimenez, Alexander Wettig, Kabir Khandpur, Yanzhe Zhang, Binyuan Hui, Ofir Press, Ludwig Schmidt, and Diyi Yang.
\newblock Swe-smith: Scaling data for software engineering agents.
\newblock \emph{arXiv preprint arXiv:2504.21798}, 2025.

\bibitem[Dong et~al.(2018)Dong, Liao, Pang, Su, Zhu, Hu, and Li]{dong2018boosting}
Yinpeng Dong, Fangzhou Liao, Tianyu Pang, Hang Su, Jun Zhu, Xiaolin Hu, and Jianguo Li.
\newblock Boosting adversarial attacks with momentum.
\newblock In \emph{Proceedings of the IEEE conference on computer vision and pattern recognition}, pages 9185--9193, 2018.

\bibitem[McCoy et~al.(2019)McCoy, Pavlick, and Linzen]{mccoy-etal-2019-right}
R~Thomas McCoy, Ellie Pavlick, and Tal Linzen.
\newblock Right for the wrong reasons: Diagnosing syntactic heuristics in natural language inference.
\newblock In \emph{Proceedings of the 57th annual meeting of the association for computational linguistics}, pages 3428--3448, 2019.

\bibitem[Singh et~al.(2025)Singh, Fry, Perelman, Tart, Ganesh, El-Kishky, McLaughlin, Low, Ostrow, Ananthram, et~al.]{singh2025openai}
Aaditya Singh, Adam Fry, Adam Perelman, Adam Tart, Adi Ganesh, Ahmed El-Kishky, Aidan McLaughlin, Aiden Low, AJ~Ostrow, Akhila Ananthram, et~al.
\newblock Openai gpt-5 system card.
\newblock \emph{arXiv preprint arXiv:2601.03267}, 2025.

\bibitem[Liu et~al.(2025)Liu, Mei, Lin, Xue, Wang, Xu, Wu, Zhang, Lin, Dong, et~al.]{liu2025deepseek}
Aixin Liu, Aoxue Mei, Bangcai Lin, Bing Xue, Bingxuan Wang, Bingzheng Xu, Bochao Wu, Bowei Zhang, Chaofan Lin, Chen Dong, et~al.
\newblock Deepseek-v3. 2: Pushing the frontier of open large language models.
\newblock \emph{arXiv preprint arXiv:2512.02556}, 2025.

\bibitem[Dihan et~al.(2026)Dihan, Awal, and Ahsan]{dihan2026patchrecall}
Mahir~Labib Dihan, Faria~Binta Awal, and Md~Ishrak Ahsan.
\newblock Patchrecall: Patch-driven retrieval for automated program repair.
\newblock \emph{arXiv preprint arXiv:2604.10481}, 2026.

\bibitem[Liang et~al.(2025)Liang, Garg, and Moghaddam]{liang2025swe}
Shanchao Liang, Spandan Garg, and Roshanak~Zilouchian Moghaddam.
\newblock The swe-bench illusion: When state-of-the-art llms remember instead of reason.
\newblock \emph{arXiv preprint arXiv:2506.12286}, 2025.

\bibitem[Ilyas et~al.(2019)Ilyas, Santurkar, Tsipras, Engstrom, Tran, and Madry]{ilyas2019adversarial}
Andrew Ilyas, Shibani Santurkar, Dimitris Tsipras, Logan Engstrom, Brandon Tran, and Aleksander Madry.
\newblock Adversarial examples are not bugs, they are features.
\newblock \emph{Advances in neural information processing systems}, 32, 2019.

\bibitem[Ribeiro et~al.(2020{\natexlab{a}})Ribeiro, Wu, Guestrin, and Singh]{ribeiro2020beyond}
Marco~Tulio Ribeiro, Tongshuang Wu, Carlos Guestrin, and Sameer Singh.
\newblock Beyond accuracy: Behavioral testing of nlp models with checklist.
\newblock In \emph{Proceedings of the 58th annual meeting of the association for computational linguistics}, pages 4902--4912, 2020{\natexlab{a}}.

\bibitem[{OpenAI}(2025)]{openai2025gpt41}
{OpenAI}.
\newblock Introducing {GPT-4.1} in the api, April 2025.
\newblock URL \url{https://openai.com/index/gpt-4-1/}.

\bibitem[{Google}(2025)]{google2026gemini31}
{Google}.
\newblock {Gemini-3.1-Pro} model card, February 2025.
\newblock URL \url{https://deepmind.google/models/model-cards/gemini-3-1-pro/}.

\bibitem[Anthropic(2026)]{claude-system-card}
Anthropic.
\newblock Claude system card, 2026.
\newblock URL \url{https://www.anthropic.com/system-cards}.

\bibitem[{Google}(2026)]{minimaxm27}
{Google}.
\newblock {MiniMax-M2.7} model card, March 2026.
\newblock URL \url{https://www.minimax.io/models/text/m27}.

\bibitem[{Qwen Team}(2026{\natexlab{a}})]{qwen_qwen3_coder_next_tech_report}
{Qwen Team}.
\newblock Qwen3-coder-next technical report.
\newblock Technical report, February 2026{\natexlab{a}}.
\newblock URL \url{https://github.com/QwenLM/Qwen3-Coder/blob/main/qwen3_coder_next_tech_report.pdf}.

\bibitem[{Qwen Team}(2026{\natexlab{b}})]{qwen3.6-35b-a3b}
{Qwen Team}.
\newblock {Qwen3.6-35B-A3B}: Agentic coding power, now open to all, April 2026{\natexlab{b}}.
\newblock URL \url{https://qwen.ai/blog?id=qwen3.6-35b-a3b}.

\bibitem[Bouzenia and Pradel(2025)]{bouzenia2025understanding}
Islem Bouzenia and Michael Pradel.
\newblock Understanding software engineering agents: A study of thought-action-result trajectories.
\newblock \emph{arXiv preprint arXiv:2506.18824}, 2025.

\bibitem[Ceka et~al.(2025)Ceka, Pujar, Ramji, Buratti, Kaiser, and Ray]{ceka2025understanding}
Ira Ceka, Saurabh Pujar, Shyam Ramji, Luca Buratti, Gail Kaiser, and Baishakhi Ray.
\newblock Understanding software engineering agents through the lens of traceability: An empirical study.
\newblock \emph{arXiv preprint arXiv:2506.08311}, 2025.

\bibitem[Kim et~al.(2026)Kim, Wang, Cui, Farmahinifarahani, Garg, Ray, Zhuo, Mukherjee, and Kumar]{kim2026trajeval}
Myeongsoo Kim, Dingmin Wang, Siwei Cui, Farima Farmahinifarahani, Shweta Garg, Baishakhi Ray, Terry~Yue Zhuo, Rajdeep Mukherjee, and Varun Kumar.
\newblock Trajeval: Decomposing code agent trajectories for fine-grained diagnosis.
\newblock \emph{arXiv preprint arXiv:2603.24631}, 2026.

\bibitem[Austin et~al.(2021)Austin, Odena, Nye, Bosma, Michalewski, Dohan, Jiang, Cai, Terry, Le, et~al.]{austin2021program}
Jacob Austin, Augustus Odena, Maxwell Nye, Maarten Bosma, Henryk Michalewski, David Dohan, Ellen Jiang, Carrie Cai, Michael Terry, Quoc Le, et~al.
\newblock Program synthesis with large language models.
\newblock \emph{arXiv preprint arXiv:2108.07732}, 2021.

\bibitem[Liu et~al.(2023{\natexlab{b}})Liu, Xia, Wang, and Zhang]{evalplus}
Jiawei Liu, Chunqiu~Steven Xia, Yuyao Wang, and Lingming Zhang.
\newblock Is your code generated by chatgpt really correct? rigorous evaluation of large language models for code generation.
\newblock \emph{Advances in neural information processing systems}, 36:\penalty0 21558--21572, 2023{\natexlab{b}}.

\bibitem[Li et~al.(2022)Li, Choi, Chung, Kushman, Schrittwieser, Leblond, Eccles, Keeling, Gimeno, Dal~Lago, et~al.]{Li_2022}
Yujia Li, David Choi, Junyoung Chung, Nate Kushman, Julian Schrittwieser, R{\'e}mi Leblond, Tom Eccles, James Keeling, Felix Gimeno, Agustin Dal~Lago, et~al.
\newblock Competition-level code generation with alphacode.
\newblock \emph{Science}, 378\penalty0 (6624):\penalty0 1092--1097, 2022.

\bibitem[Lai et~al.(2023)Lai, Li, Wang, Zhang, Zhong, Zettlemoyer, Yih, Fried, Wang, and Yu]{Lai2022DS1000}
Yuhang Lai, Chengxi Li, Yiming Wang, Tianyi Zhang, Ruiqi Zhong, Luke Zettlemoyer, Wen-tau Yih, Daniel Fried, Sida Wang, and Tao Yu.
\newblock Ds-1000: A natural and reliable benchmark for data science code generation.
\newblock In \emph{International Conference on Machine Learning}, pages 18319--18345. PMLR, 2023.

\bibitem[Gu et~al.(2024)Gu, Rozi{\`e}re, Leather, Solar-Lezama, Synnaeve, and Wang]{gu2024cruxeval}
Alex Gu, Baptiste Rozi{\`e}re, Hugh Leather, Armando Solar-Lezama, Gabriel Synnaeve, and Sida~I Wang.
\newblock Cruxeval: A benchmark for code reasoning, understanding and execution.
\newblock \emph{arXiv preprint arXiv:2401.03065}, 2024.

\bibitem[Zhang et~al.(2023)Zhang, Chen, Zhang, Liu, Zan, Mao, Lou, and Chen]{zhang2023repocoder}
Fengji Zhang, Bei Chen, Yue Zhang, Jin Liu, Daoguang Zan, Yi~Mao, Jian-Guang Lou, and Weizhu Chen.
\newblock Repocoder: Repository-level code completion through iterative retrieval and generation.
\newblock \emph{arXiv preprint arXiv:2303.12570}, 2023.

\bibitem[Ding et~al.(2023)Ding, Wang, Ahmad, Ding, Tan, Jain, Ramanathan, Nallapati, Bhatia, Roth, et~al.]{ding2023crosscodeeval}
Yangruibo Ding, Zijian Wang, Wasi Ahmad, Hantian Ding, Ming Tan, Nihal Jain, Murali~Krishna Ramanathan, Ramesh Nallapati, Parminder Bhatia, Dan Roth, et~al.
\newblock Crosscodeeval: A diverse and multilingual benchmark for cross-file code completion.
\newblock \emph{Advances in Neural Information Processing Systems}, 36:\penalty0 46701--46723, 2023.

\bibitem[Wang et~al.(2025)Wang, Ramalho, Celestino, Pham, Liu, Sinha, Portillo, Osunwa, and Maduekwe]{wang2025swebenchpp}
Lilin Wang, Lucas Ramalho, Alan Celestino, Phuc~Anthony Pham, Yu~Liu, Umang~Kumar Sinha, Andres Portillo, Onassis Osunwa, and Gabriel Maduekwe.
\newblock Swe-bench++: A framework for the scalable generation of software engineering benchmarks from open-source repositories.
\newblock \emph{arXiv preprint arXiv:2512.17419}, 2025.

\bibitem[Taori et~al.(2020)Taori, Dave, Shankar, Carlini, Recht, and Schmidt]{10.5555/3495724.3497285}
Rohan Taori, Achal Dave, Vaishaal Shankar, Nicholas Carlini, Benjamin Recht, and Ludwig Schmidt.
\newblock Measuring robustness to natural distribution shifts in image classification.
\newblock \emph{Advances in Neural Information Processing Systems}, 33:\penalty0 18583--18599, 2020.

\bibitem[Szegedy et~al.(2013)Szegedy, Zaremba, Sutskever, Bruna, Erhan, Goodfellow, and Fergus]{szegedy2014intriguingproperties}
Christian Szegedy, Wojciech Zaremba, Ilya Sutskever, Joan Bruna, Dumitru Erhan, Ian Goodfellow, and Rob Fergus.
\newblock Intriguing properties of neural networks.
\newblock \emph{arXiv preprint arXiv:1312.6199}, 2013.

\bibitem[Goodfellow et~al.(2014)Goodfellow, Shlens, and Szegedy]{goodfellow2015explainingharness}
Ian~J Goodfellow, Jonathon Shlens, and Christian Szegedy.
\newblock Explaining and harnessing adversarial examples.
\newblock \emph{arXiv preprint arXiv:1412.6572}, 2014.

\bibitem[Ribeiro et~al.(2020{\natexlab{b}})Ribeiro, Wu, Guestrin, and Singh]{ribeiro-etal-2020-beyond}
Marco~Tulio Ribeiro, Tongshuang Wu, Carlos Guestrin, and Sameer Singh.
\newblock Beyond accuracy: Behavioral testing of nlp models with checklist.
\newblock In \emph{Proceedings of the 58th annual meeting of the association for computational linguistics}, pages 4902--4912, 2020{\natexlab{b}}.

\bibitem[Gardner et~al.(2020)Gardner, Artzi, Basmov, Berant, Bogin, Chen, Dasigi, Dua, Elazar, Gottumukkala, et~al.]{gardner-etal-2020-evaluating}
Matt Gardner, Yoav Artzi, Victoria Basmov, Jonathan Berant, Ben Bogin, Sihao Chen, Pradeep Dasigi, Dheeru Dua, Yanai Elazar, Ananth Gottumukkala, et~al.
\newblock Evaluating models’ local decision boundaries via contrast sets.
\newblock In \emph{Findings of the Association for Computational Linguistics: EMNLP 2020}, pages 1307--1323, 2020.

\bibitem[Wang et~al.(2023)Wang, Li, Qian, Yang, Wang, Shang, Kumar, Tan, Ray, Bhatia, et~al.]{wang-etal-2023-recode}
Shiqi Wang, Zheng Li, Haifeng Qian, Chenghao Yang, Zijian Wang, Mingyue Shang, Varun Kumar, Samson Tan, Baishakhi Ray, Parminder Bhatia, et~al.
\newblock Recode: Robustness evaluation of code generation models.
\newblock In \emph{Proceedings of the 61st Annual Meeting of the Association for Computational Linguistics (Volume 1: Long Papers)}, pages 13818--13843, 2023.

\bibitem[Wen et~al.(2025)Wen, Hu, Guo, Cordy, and Traon]{variablerenaming2025}
Jin Wen, Qiang Hu, Yuejun Guo, Maxime Cordy, and Yves~Le Traon.
\newblock Variable renaming-based adversarial test generation for code model: Benchmark and enhancement.
\newblock \emph{ACM Transactions on Software Engineering and Methodology}, 35\penalty0 (1):\penalty0 1--28, 2025.

\bibitem[Li et~al.(2023)Li, Wang, Liu, Wang, Chen, Wang, and Gao]{li2023cctesttestingrepairingcode}
Zongjie Li, Chaozheng Wang, Zhibo Liu, Haoxuan Wang, Dong Chen, Shuai Wang, and Cuiyun Gao.
\newblock Cctest: Testing and repairing code completion systems.
\newblock In \emph{2023 IEEE/ACM 45th International Conference on Software Engineering (ICSE)}, pages 1238--1250. IEEE, 2023.

\bibitem[Na et~al.(2023)Na, Choi, and Lee]{na-etal-2023-dip}
CheolWon Na, YunSeok Choi, and Jee-Hyong Lee.
\newblock Dip: Dead code insertion based black-box attack for programming language model.
\newblock In \emph{Proceedings of the 61st Annual Meeting of the Association for Computational Linguistics (Volume 1: Long Papers)}, pages 7777--7791, 2023.

\bibitem[Orvalho and Kwiatkowska(2025)]{orvalho2025largelanguagemodelsrobust}
Pedro Orvalho and Marta Kwiatkowska.
\newblock Are large language models robust in understanding code against semantics-preserving mutations?
\newblock \emph{arXiv preprint arXiv:2505.10443}, 2025.

\bibitem[Maveli et~al.(2025)Maveli, Vergari, and Cohen]{maveli2025largelanguagemodelscapture}
Nickil Maveli, Antonio Vergari, and Shay~B Cohen.
\newblock What can large language models capture about code functional equivalence?
\newblock In \emph{Findings of the Association for Computational Linguistics: NAACL 2025}, pages 6865--6903, 2025.

\bibitem[Lam et~al.(2025)Lam, Wang, Huang, and Lyu]{lam2025codecrash}
Man~Ho Lam, Chaozheng Wang, Jen{-}tse Huang, and Michael~R. Lyu.
\newblock Codecrash: Stress testing llm reasoning under structural and semantic perturbations.
\newblock \emph{arXiv preprint arXiv:2504.14119}, 2025.

\bibitem[Lee et~al.(2025)Lee, Kim, Singh, Pereira, Sonwane, White, Stengel-Eskin, Bansal, Shi, Sordoni, et~al.]{lee2025gistifycodebaselevelunderstandingruntime}
Hyunji Lee, Minseon Kim, Chinmay Singh, Matheus Pereira, Atharv Sonwane, Isadora White, Elias Stengel-Eskin, Mohit Bansal, Zhengyan Shi, Alessandro Sordoni, et~al.
\newblock Gistify! codebase-level understanding via runtime execution.
\newblock \emph{arXiv preprint arXiv:2510.26790}, 2025.

\bibitem[Kwon et~al.(2023)Kwon, Li, Zhuang, Sheng, Zheng, Yu, Gonzalez, Zhang, and Stoica]{kwon2023efficient}
Woosuk Kwon, Zhuohan Li, Siyuan Zhuang, Ying Sheng, Lianmin Zheng, Cody~Hao Yu, Joseph~E. Gonzalez, Hao Zhang, and Ion Stoica.
\newblock Efficient memory management for large language model serving with pagedattention.
\newblock In \emph{Proceedings of the ACM SIGOPS 29th Symposium on Operating Systems Principles}, 2023.

\end{thebibliography}

\newpage
\appendix



\section{Experimental Details}
\label{sec:experimental-details}

\subsection{Compute Resources}
\label{sec:compute_resouces}

All experiments were executed in Docker-based SWE-Bench repository environments on Ubuntu 20.04.6 LTS servers. The main execution server was equipped with two AMD EPYC 7542 32-core processors, approximately 1 TiB RAM, and 6.8 TB storage, using Docker 27.1.1. Most proprietary models were accessed through their commercial APIs, so no local GPU inference was required for these models. For open-weight models, Qwen3-Coder-Next and Qwen3.6-35B-A3B were deployed locally using vLLM~\citep{kwon2023efficient} on four NVIDIA A800-SXM4-80GB GPUs.

\subsection{Agent Framework and Execution Protocol}
\label{sec:execution_protocol}

To ensure fair comparison across models, we use the same agent framework and execution protocol for all experiments. Each evaluated model is instantiated as the backbone of \texttt{mini-swe-agent}~\citep{yang2024swe} and runs under the standard bash-only setting. The agent interacts with the repository through shell commands. We keep the agent framework, prompt format, execution environment, and stopping criteria fixed across all models and task settings. The only varying component is the backbone language model.

For each instance, the agent is initialized in the corresponding repository environment using docker images. The input includes the task instruction and the current repository state. The agent proceeds until it submits a final answer, reaches the maximum step budget of 250, which is the same as the standard bash-only setting. Unless otherwise specified, we use default temperature, top-p, and maximum generation length. 

\subsection{Models}

We evaluate eight frontier language models as agent backbones: GPT-4.1, GPT-5, DeepSeek-V3.2, Gemini-3.1-Pro, Claude-Sonnet-4.6, MiniMax-M2.7, Qwen3-Coder-Next, and Qwen3.6-35B-A3B. These models are selected to cover major model families from different providers, including both proprietary and open-weight systems. For each family, we choose either the latest available model or a widely used stable version at the time of our experiments. This selection is intended to provide broad and representative coverage of current code-agent backbones, rather than focusing on a single model provider or model type.

\subsection{Perturbation Construction and Functionality Preservation}
\label{sec:perturb_construction}

We implement repository perturbations mainly with LibCST. Instead of relying on unstable regular-expression rewriting, we transform source files use the syntax-tree. This allows us to preserve the original code structure more reliably and to ensure that the extracted dependency paths, externalized values and other expected structures follow our intended perturbation design. We validate each perturbed instance from two perspectives, functionality preservation and solvability. For functionality preservation, after applying \textsc{RepoMirage}-Perturb to the original repository, we run the official SWE-Bench \texttt{PASS\_TO\_PASS} test scripts to ensure that existing correct behaviors are not broken by our transformations. For solvability, we first apply the gold patch to the original repository, then apply the same perturbations, and finally run the official \texttt{FAIL\_TO\_PASS} test scripts. This verifies that the perturbed instance remains solvable under the gold solution and that the original task are preserved.

For \textbf{Dependency-path indirection}, we only construct proxy chains for absolute imports whose targets are independent of the repository-internal file layout. For example, modules such as \texttt{json} and \texttt{re} can be safely rerouted because they do not depend on relative paths inside the benchmark repository. This avoids introducing ambiguity or import errors caused by fragile relative-import semantics. For each imported module, the generated proxy-chain path is uniquely determined, so the dependency routing is deterministic and reproducible.

For \textbf{Runtime-target masking}, we rename the original target file and create a package-level wrapper using the original target-file name. This preserves the original import path, so other modules in the repository can still reference the target through the same package name after perturbation. It also keeps the task partially solvability for issues that can only be traced through the package name. An agent may still locate the wrapper, but must further inspect its loading logic to identify the actual runtime implementation file. Instead of relying on direct relative imports, the wrapper dynamically loads the renamed implementation file from its explicit file path while preserving the intended package name. This avoids edge cases and ensures that the runtime binding remains consistent after the perturbation. Meanwhile, the generated fake files are placed near the real runtime file, but they are never imported or executed by the program. They therefore do not affect repository functionality.

For \textbf{Local-value externalization}, we extract only local constant values from the target file. When more than three constants are extracted, we group every three values into one external JSON file, increasing the number of cross-file dependencies. The target file reads all related JSON files, merges them into a single dictionary, and replaces the original values with corresponding dictionary-key references. This transformation preserves the logic while moving value definitions outside the file.

Overall, the three perturbations mainly modify files around the target file and the relations between the target file and external resources. They only introduce little change to the internal control flow and semantic logic of the original code. This design allows \textsc{RepoMirage} to focus on repository context reasoning, rather than evaluating robustness to code obfuscation.

\subsection{Task Construction and Validation}
\label{sec:task_construction}

For \textsc{RepoMirage}-Extend, we also construct tasks under two perspectives, solvability and verifiability. A valid task should have at least one feasible solution, and the submitted solution should be automatically checkable by deterministic scripts. All \textsc{RepoMirage} tasks are built on top of \textsc{RepoMirage}-Perturb instances. That is, each task is constructed in a repository where all three perturbations have already been applied. Agents still interact with the repository through the standard \texttt{mini-swe-agent} command-line interface.

Before creating each derived task, we first run the corresponding checker on \textsc{RepoMirage}-Perturb repository before erasing proxy files, modifying wrappers, or removing dependency keys. The checker must pass at this stage, which ensures that the task has at least one known feasible solution. During evaluation, the task prompt specifies the exact set of files that the agent is allowed to modify. The checker also verifies that no irrelevant files are changed, preventing agents from bypassing the intended task objective through unrelated edits.

For \textbf{Proxy Chain Completion}, we erase proxy files located in the middle of dependency-routing paths. The erased proxy files are selected so that their upstream and downstream nodes remain intact, allowing the missing forwarding logic to be inferred from surrounding files. We also select erased proxy files from different module-routing paths to avoid reducing the task to repeated copying. Ideally, the agent should first locate the erased proxy files, trace which lower-level proxy files import them, follow the chain to the runtime target file, infer which real module each chain is intended to expose, and then restore the erased forwarding files accordingly. For validation, the agent is only allowed to modify the erased proxy files. After applying the agent patch, the checker imports the dependency aliases exposed from the target file and compares the resolved modules with the expected original modules. In this way, we evaluate whether the dependency path has been functionally restored, without requiring the agent to reproduce the exact original proxy implementation.

For \textbf{Runtime Target Identification}, the task is constructed by removing the wrapper reference to the real runtime file. The main challenge is to distinguish the actual implementation file from nearby fake files. These fake files are generated through small but behaviorally meaningful changes, such as boolean-condition reversal, shuffled import aliases, or mismatched dependency-key usage. Such files remain superficially similar to the real target file, but can be distinguished by reasoning about repository context and by designing targeted runtime probes. Ideally, the agent should identify the wrapper, collect the candidate implementation files, understand their roles in the local directory, and determine which file should be loaded by the wrapper. For validation, the agent is only allowed to modify the wrapper \texttt{\_\_init\_\_.py}. The checker then verifies whether the wrapper correctly resolves to the real runtime implementation file.

For \textbf{Missing Constant Recovery}, we erase selected keys in external dependency JSON files while keeping their values unchanged. Ideally, the agent should first locate the JSON files with missing keys, inspect the target file that loads and merges these JSON resources, identify where the missing keys are referenced, and infer the role of each missing value from the surrounding code context, such as loop control, index offsets, or special arithmetic constants. The agent then matches the erased keys with the remaining values. For validation, the agent is only allowed to modify the JSON files containing erased keys. The checker loads all related JSON files and merges them into a dictionary following the same logic as the target file, then verifies whether the reconstructed key-value relations are correct. This task does not require a unique file-level assignment: if equivalent values appear in different JSON files and are merged into the same final dictionary, the agent only needs to recover the correct key-value mapping after merging.

For \textbf{Multi-File Issue Resolution}, the tasks are selected from SWE-Bench Verified instances whose gold patches modify multiple files. Since these tasks are still issue-resolution tasks on perturbed SWE-Bench repositories, their solvability and validation follow the original SWE-Bench protocol. The agent must submit a patch that passes the official evaluation tests.

Overall, \textsc{RepoMirage}-Extend converts the structural bottlenecks introduced by \textsc{RepoMirage}-Perturb into solvable and automatically verifiable tasks. These tasks can be generated through rule-based procedures and checked with deterministic scripts. By making the perturbed repository relations explicit task objectives, \textsc{RepoMirage}-Extend amplifies the evaluation signal for repository context reasoning.

\subsection{File Access and Trajectory Metrics}

We record the full shell-command trajectory generated by \texttt{mini-swe-agent} for each run. To analyze agent behavior, we first classify each command into four operation types: \textit{edit}, \textit{test}, \textit{search}, and \textit{read}. The classification is implemented by a rule-based matcher over shell commands. Editing commands include patch application, in-place rewriting, file creation, file deletion, and script-based file writes. Testing commands include unit-test execution, task-specific checkers, compilation checks, and runtime validation commands. Search commands include repository-level locating operations such as \texttt{rg}, \texttt{grep}, \texttt{find}, \texttt{ls}, and directory traversal. Read commands include commands that inspect file contents or metadata, such as \texttt{cat}, \texttt{sed -n}, \texttt{head}, \texttt{tail}, and script-based file reads.

For transition analysis, we merge \textit{search} and \textit{read} into a single \textit{explore} state, because both correspond to gathering repository information before or during problem solving. This produces a three-state trajectory and we compute the transition probability use the formula in Section~\ref{sec:exploration-metrics}. We compute transition matrices for each trajectory and aggregate them across instances for each model and task setting. When comparing two settings, we report the difference in transition probabilities, denoted as $\Delta p_{a \rightarrow b}$.

We also compute file-access statistics from the same trajectories. A repository file is counted as accessed only when its contents are explicitly exposed to the agent, mainly through read commands or content-producing search commands. We normalize file paths and deduplicate repeated accesses to the same file within one trajectory. Directory listing commands, such as \texttt{ls} and \texttt{find}, are treated as exploration actions, but files merely listed by these commands are not counted as accessed unless their contents are later inspected. Temporary files, generated logs, patch files, virtual-environment files, and version-control metadata are excluded from the file-access count. Finally, we measure the exploration stage before editing. The exploration-stage length is defined as the share of steps before the first edit operation. These metrics indicate whether agents can convert repository exploration into concrete editing decisions, or instead remain in repeated exploration under stronger repository-context demands.

\begin{figure*}[t]
  \centering
  \includegraphics[width=1.0 \columnwidth]{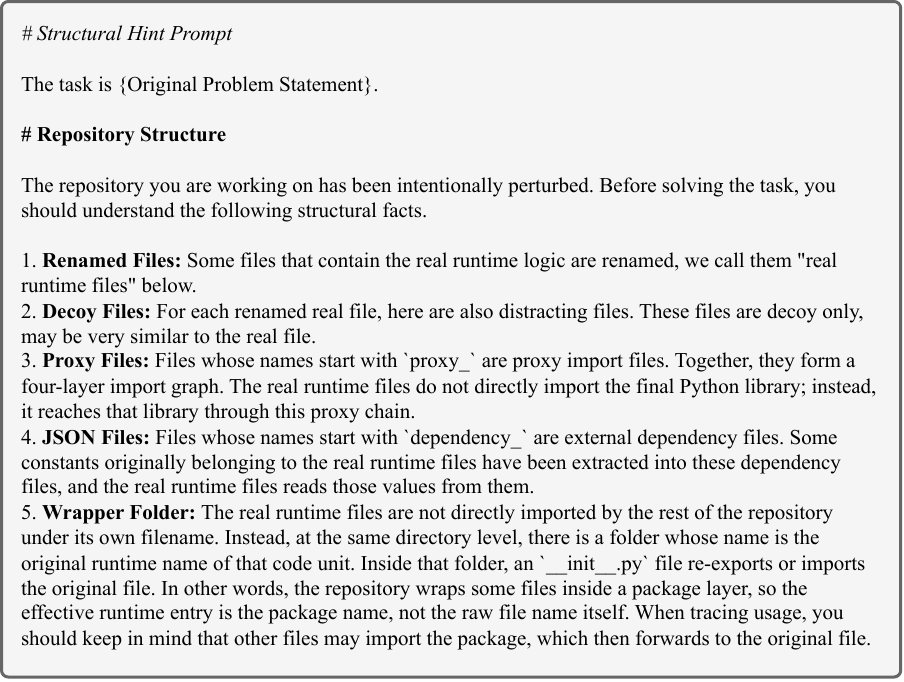}
  \caption{Prompt for repository structural hints.}
  \label{fig:hint_prompt}
\end{figure*}

\subsection{Structural Hint Prompts}
\label{sec:hint_details}

To diagnose whether agents fail because they cannot access repository information or because they cannot organize it into a useful structural understanding, we additionally evaluate a hint-based setting. In this setting, we prepend a structural hint prompt to the original task instruction. The hint describes the general perturbation mechanisms used in \textsc{RepoMirage}-Perturb, including renamed runtime files, decoy files, proxy import chains, external dependency JSON files, and wrapper folders. An example prompt is shown in Fig.~\ref{fig:hint_prompt}.

The structural hints do not reveal the concrete solution of an instance. They do not specify which file should be edited, which proxy file should be restored, which candidate file is the real runtime target, or which missing key should be filled. Instead, they only describe the repository-level structural facts that are shared across perturbed instances. Therefore, the agent still needs to inspect the repository, trace file relations, and produce the final solution by itself.

We use the same hint template for all models under the corresponding task setting. The only instance-specific part is the original problem statement, which is inserted into the prompt. This design makes the hint experiment a controlled diagnostic intervention. It approximates the structural understanding that an ideal exploration process should recover, while keeping the downstream reasoning and editing process unchanged.

\begin{figure*}[t]
  \centering
  \includegraphics[width=1.0 \columnwidth]{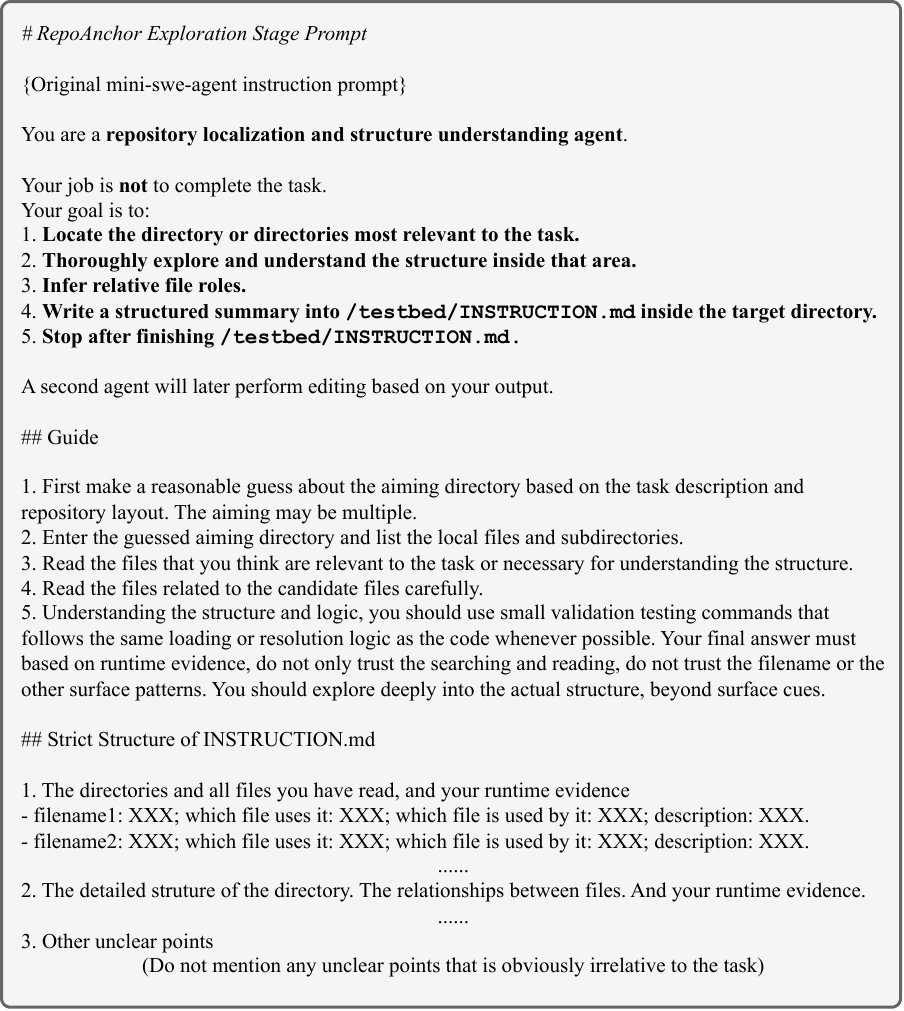}
  \caption{Prompt for exploration stage of \textsc{RepoAnchor}.}
  \label{fig:repoanchor_exploration_prompt}
\end{figure*}

\begin{figure*}[t]
  \centering
  \includegraphics[width=1.0 \columnwidth]{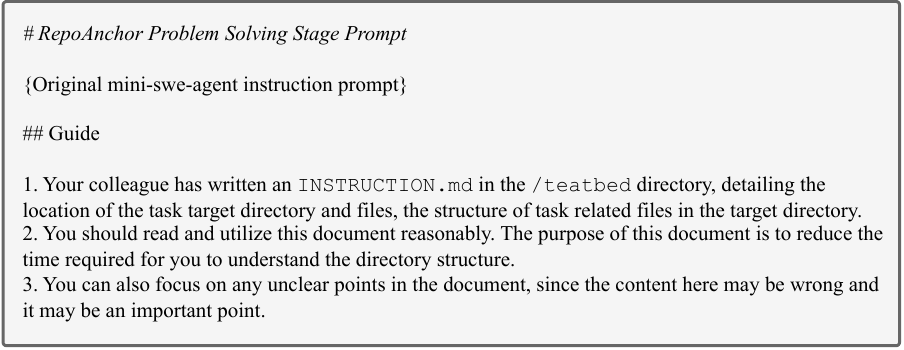}
  \caption{Prompt for problem solving stage of \textsc{RepoAnchor}.}
  \label{fig:repoanchor_exploration_solving}
\end{figure*}

\subsection{\textsc{RepoAnchor} Prompts}

RepoAnchor uses a two-stage prompting strategy to separate repository-structure exploration from downstream task solving. In the first stage, the agent works for repository localization and structure-understanding. As shown in Fig.~\ref{fig:repoanchor_exploration_prompt}, this agent is explicitly instructed not to complete the task or edit source code. Instead, it must identify the task-relevant directory, explore the local file structure, infer file roles and dependency relations, and write the resulting structural summary into \texttt{INSTRUCTION.md}.

The exploration prompt emphasizes runtime evidence rather than surface-level cues. The agent is encouraged to inspect files, trace import or loading relations, and use lightweight validation commands when necessary. The generated \texttt{INSTRUCTION.md} follows a fixed structure, including the files inspected, their roles, relationships between files, runtime evidence, and remaining uncertainties. This intermediate document serves as the structural anchor passed to the second-stage agent.

In the second stage, the agent also uses an additional guide to read and utilize the generated \texttt{INSTRUCTION.md}, as shown in Fig.~\ref{fig:repoanchor_exploration_solving}. The prompt reminds the agent that the summary contains the localized target directory, relevant files, and structural relations produced by the exploration-stage agent. It also allows the agent to verify uncertain points in the summary rather than blindly following it. Thus, \textsc{RepoAnchor} separates repository exploration from code construction: the first stage produces an explicit structural anchor, and the second stage uses it to guide editing, testing, and submission.

\begin{figure*}[t]
  \centering
  \includegraphics[width=1.0 \columnwidth]{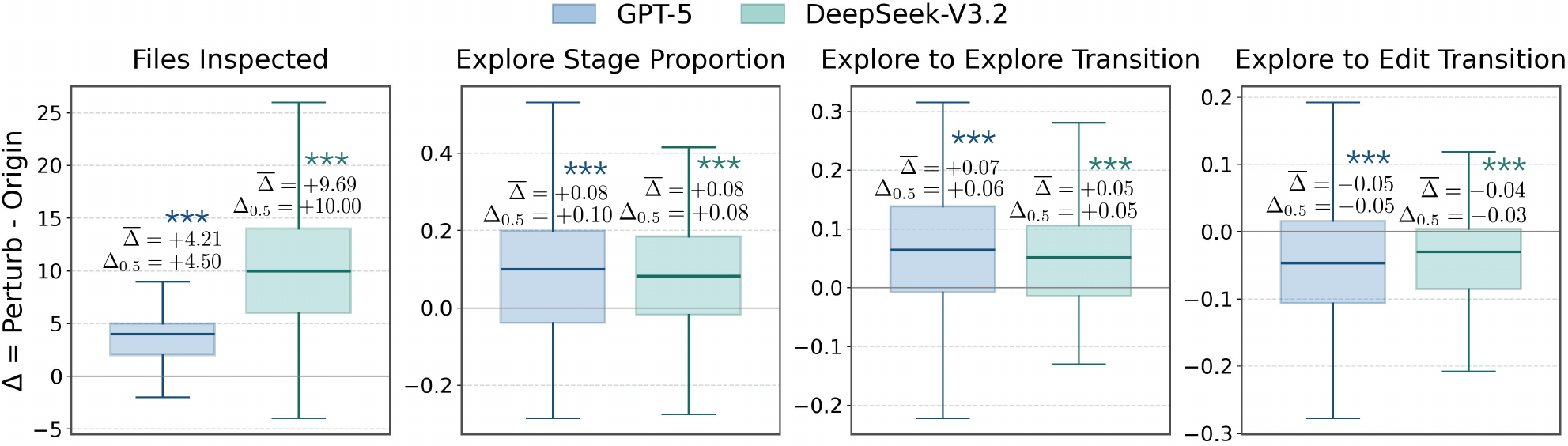}
  \caption{\textbf{Behavior shifts under \textsc{RepoMirage}-Perturb.} $\Delta$ measures the change from SWE-Bench Verified to \textsc{RepoMirage}-Perturb. Files Inspected counts distinct opened files; Explore Stage Proportion is the pre-edit step ratio; transition metrics report action-transition changes. $\bar{\Delta}$ and $\Delta_{0.5}$ denote the mean and median, respectively.}
  \label{fig:perturb_metric}
\end{figure*}

\begin{figure*}[t]
  \centering
  \includegraphics[width=1.0 \columnwidth]{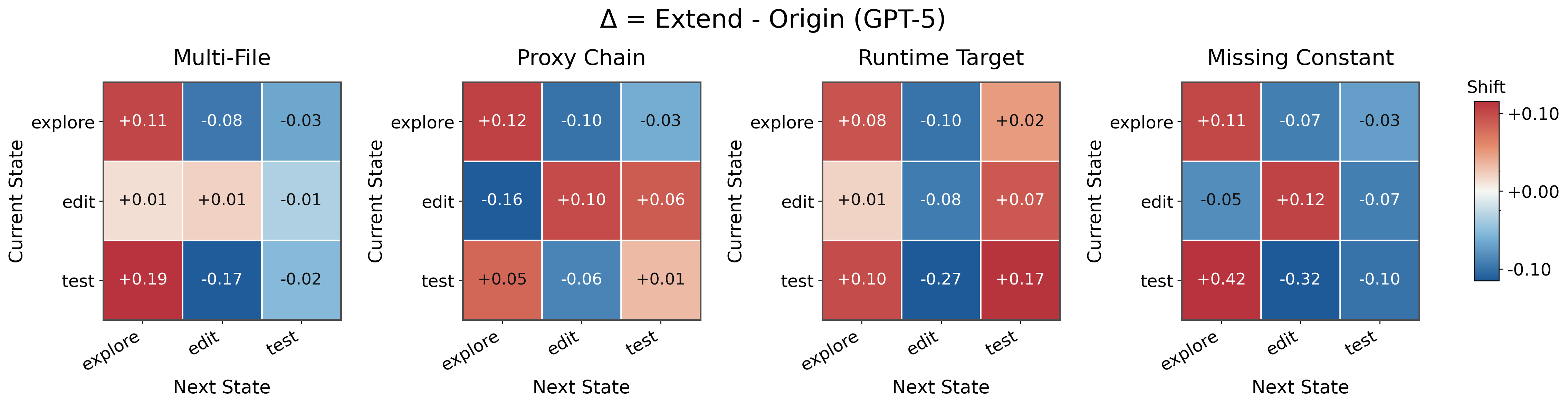}
  \caption{\textbf{Task-level transition probability differences.} Each heatmap corresponds to one \textsc{RepoMirage}-Extend task for GPT-5. Rows denote the current action state, columns denote the next action state, and each cell reports the transition-probability difference between the \textsc{RepoMirage}-Extend setting and SWE-Bench setting. Red cells indicate positive shifts, blue cells indicate negative shifts, and the annotated values give the exact differences.}
  \label{fig:task_analysis_heatmaps}
\end{figure*}

\section{Additional Analysis and Examples}

\subsection{Behavior Shifts on \textsc{RepoMirage}-Perturb}
\label{sec:traj_on_perturb}

~\cref{fig:perturb_metric} shows the behavior shifts on \textsc{RepoMirage}-Perturb, where agents still solve the original issue-resolution task but under perturbed repository structures. The results show a pattern consistent with the trajectory changes observed on \textsc{RepoMirage}-Extend. After perturbation, both GPT-5 and DeepSeek-V3.2 inspect more files and spend a larger proportion of their trajectories before the first edit operation, indicating that the perturbations indeed increase the need for repository exploration. At the transition level, the probability of exploration-to-exploration transitions increases, while the probability of exploration-to-edit transitions decreases. This suggests that even when the task objective remains unchanged, repository-level perturbations make agents more likely to stay in repeated exploration rather than turning gathered information into concrete edits. These results further support the main paper's observation that stronger repository-context demands induce exploration drift.

\subsection{Task-Level Transition Comparison}
\label{sec:transition_analysis}

~\cref{fig:task_analysis_heatmaps} provides task-level transition heatmaps for GPT-5 on \textsc{RepoMirage}-Extend. We observe a consistent pattern across all four tasks: the transition from exploration to exploration increases, while the transition from exploration to editing decreases. This consistent pattern further supports the observation in the main paper that agents tend to remain in repeated exploration under stronger repository-context demands, rather than converting the gathered information into concrete editing actions.

At the same time, the heatmaps also reveal task-specific differences. In \textbf{Multi-File Issue Resolution}, the increase from testing back to exploration suggests that failed tests often expose missing related files, pushing the agent to search for additional edit locations. In \textbf{Proxy Chain Completion}, the strong exploration persistence reflects the need to trace dependency paths across several proxy files before making a local restoration. In \textbf{Runtime Target Identification}, the larger testing-related transitions suggest that agents often rely on runtime probes to distinguish the real target file from decoys. In \textbf{Missing Constant Recovery}, the sharp shift from testing back to exploration indicates that failed validation frequently forces the agent to revisit the relation between JSON values and their usage contexts. These differences show that the four tasks impose distinct forms of repository-context reasoning, while sharing the same general exploration bottleneck.

\subsection{Example of \textsc{RepoAnchor} Instruction Summary}

\begin{figure*}[t]
  \centering
  \includegraphics[width=1.0 \columnwidth]{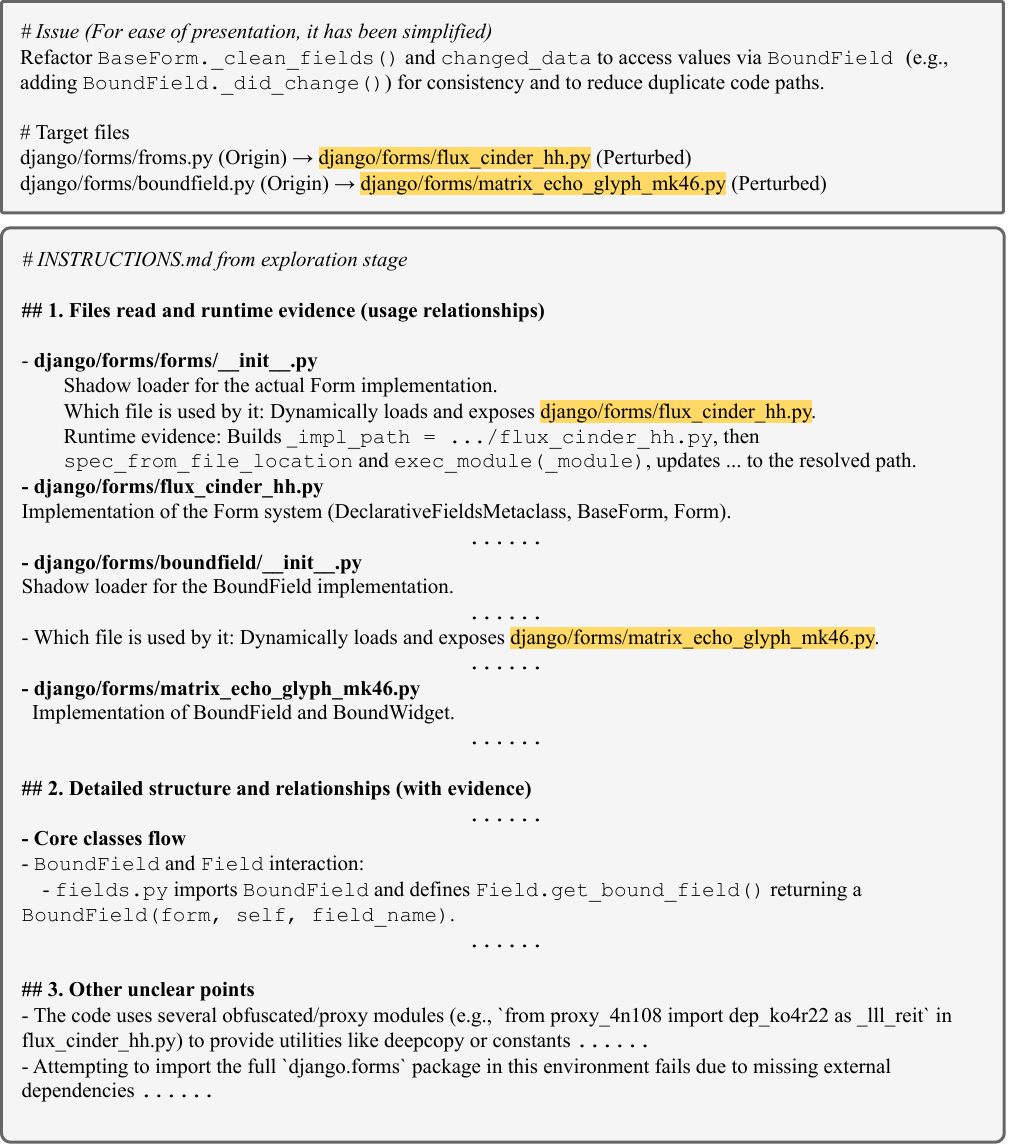}
  \caption{An example of \texttt{INSTRUCTION.md} generated by \textsc{RepoAnchor}}
  \label{fig:repoanchor_example}
\end{figure*}

~\cref{fig:repoanchor_example} shows an example of the intermediate \texttt{INSTRUCTIONS.md} generated by \textsc{RepoAnchor} on the instance \texttt{django\_\_django-14631} with GPT-5. In the exploration stage, \textsc{RepoAnchor} identifies the wrapper modules, the renamed runtime files, and their usage relationships. For example, it records that \texttt{django/forms/\_\_init\_\_.py} dynamically loads the perturbed implementation file for the form system, while \texttt{django/forms/boundfield/\_\_init\_\_.py} loads the perturbed implementation file for \texttt{BoundField}. It also summarizes the core relation between \texttt{Field.get\_bound\_field()} and \texttt{BoundField}, which is directly relevant to the issue.

This example illustrates how \textsc{RepoAnchor} converts scattered repository observations into an explicit structural summary. Instead of forcing the problem-solving agent to rediscover wrappers, renamed files, and cross-file relations from scratch, the generated \texttt{INSTRUCTION.md} provides a compact anchor for later editing and validation.

\section{Further Discussions}

\subsection{Limitations}
\label{sec:limitations}

\textsc{RepoMirage} still has several limitations. First, due to the substantial engineering complexity required by repository-level perturbations, our current construction is limited to Python repositories, where syntax-tree transformation and runtime validation are relatively tractable. Extending \textsc{RepoMirage} to languages with more complex build systems or stricter compilation constraints remains future work. Second, our tasks still follow a relatively classical problem-solving format, where agents are asked to modify a repository to satisfy a given objective. Future evaluations could further explore more diverse software-engineering scenarios, such as code review, repository-level test generation, refactoring, and maintenance-oriented tasks.

\subsection{Broader Impacts}
\label{sec:broader_impact}

\textsc{RepoMirage} can have positive impacts by helping researchers and practitioners better evaluate the reliability of code agents, reducing over-reliance on benchmark scores that may not reflect genuine repository context reasoning. At the same time, stronger repository-aware agents may increase the capability of automated code modification systems, which could be misused for harmful code changes or over-trusted in high-stakes software maintenance. We therefore encourage using \textsc{RepoMirage} as a diagnostic benchmark together with controlled deployment, human review, and repository-level validation safeguards.

\section{Release Format and Reproducibility}
\label{sec:release}

We provide an anonymized open-source repository at 
\url{https://anonymous.4open.science/r/RepoMirage}. 
\textsc{RepoMirage} is released as an executable benchmark-generation and evaluation toolkit rather than a separately hosted static dataset. We do not redistribute modified copies of SWE-Bench repositories or Docker images. Instead, our code operates on existing SWE-Bench-compatible Docker environments and reconstructs perturbed repositories and derived task environments through documented scripts.

The released code contains two main components. \texttt{RepoMirage\_Perturb/} builds \textsc{RepoMirage}-Perturb repository images, applies repository-level perturbations, and exports per-instance metadata. \texttt{RepoMirage\_Extend/} uses this metadata to assign instances to task families, generate \textsc{RepoMirage}-Extend task images, and provide deterministic validation scripts for agent submissions. The repository also includes optional utilities for exporting generated task lists into local Hugging Face-style datasets for mini-swe-agent-style runners.

Users can reproduce the benchmark by first preparing the original SWE-Bench Verified environments, then running the released perturbation and task-construction workflows. This release format keeps the source benchmark assets under their original distribution channels while making our perturbation procedures, task-generation rules, checkers, and evaluation logic reproducible. Since \textsc{RepoMirage} does not introduce a separately hosted static dataset, we provide executable code and references to the SWE-Bench Verified resources instead of a dataset metadata file.

\section{The Use of Large Language Models}
\label{sec:llm}

We used large language models (LLMs) only to assist with polishing the writing of this paper, including grammar checking and rephrasing for clarity. No LLM was used to generate research ideas, design experiments, conduct analysis, produce results, or write substantive technical content. All technical contributions, analyses, and results are the authors' own work.



\end{document}